\documentclass[aps,twocolumn,superscriptaddress]{revtex4}
\usepackage{graphics} 
\usepackage{epsfig} 
\usepackage{amsmath}
\usepackage{amsfonts}
\usepackage{color}

\begin{document}
\title{Global linear stability analysis of kinetic Trapped Ion Mode (TIM) 
in tokamak plasma using spectral method}
\author{D. Mandal}
\author{M. Lesur}
\author{E. Gravier}
\author{J. N. Sama}
\author{A. Guillevic}
\affiliation{Universit\'{e} de Lorraine, CNRS, IJL, F-54000 Nancy, France}
\author{Y. Sarazin}
\author{X. Garbet}
\affiliation{Association Euratom-CEA, CEA/DSM/DRFC, Centre de Cadarache,\\ F-13108, St-Paul-Lez-Durance, France}

\begin{abstract}
Trapped ion modes (TIM)  belong to the family of ion temperature gradient
(ITG) modes, which are one of the important ingredients in heat turbulent
transport at the ion scale in tokamak plasmas. 
A global linear analysis of a reduced gyro-bounce kinetic model for trapped 
particle modes is performed, and a spectral method is proposed to solve the 
dispersion relation. Importantly, the radial profile of the particle drift 
velocity is taken into account in the linear analysis by considering the 
magnetic flux $\psi$ dependency of the 
equilibrium Hamiltonian $H_{eq}(\psi)$ in the 
quasi-neutrality equation and equilibrium gyro-bounce averaged distribution 
function $F_{eq}$. Using this spectral method,  
linear growth-rates of TIM instability in presence of different temperature
profiles and precession frequencies of trapped ions, with 
an approximated constant Hamiltonian and the exact $\psi$ 
dependent equilibrium Hamiltonian, are investigated. 
The growth-rate depends on the logarithmic gradient of temperature 
$\kappa_T$, density $\kappa_n$ and equilibrium Hamiltonian 
$\kappa_{\Lambda}$. With the exact $\psi$ dependent Hamiltonian, 
the growth rates and potential profiles are modified significantly, 
compared to the cases with approximated constant Hamiltonian. 
All the results from the global linear analysis agree with a 
semi-Lagrangian based linear Vlasov solver with a good accuracy.
This spectral method is very fast and 
requires very less computation resources compared to a linear version of 
Vlasov-solver based on a semi-Lagrangian scheme. 
\end{abstract}



\maketitle
\section{Introduction \label{introduction}}
Low frequency and low wavenumber turbulence, which is mainly generated by ion 
temperature gradient (ITG) \cite{dimits01:a, guzdar82:p, ottaviani97:m} and 
trapped electron mode (TEM) \cite{stallard:b, dannert:t} instabilities, plays a
dominant role in the anomalous radial energy and particle transport in 
magnetically confined fusion plasmas. ITG is an important ingredient 
in anomalous ion heat transport in tokamak whereas 
TEM turbulence drives electron particle and heat transport. 
Ion temperature gradient driven modes are 
frequently observed and relevant in tokamak plasma experiments 
\cite{brower:d,eubank:h}. Trapped ion mode (TIM) belongs 
to this family of ion temperature gradient modes, and is driven by the resonant 
motion of trapped ions \cite{depret00:g}. The trapped ion 
instability is characterized by frequencies of the order of the trapped ion 
precession frequency and radial scales of the order of several banana widths.
Though TIM modes have lower frequency and long wavelength compared to TEM modes,
TIM has similar physical mechanisms as TEM in the linear regime. 
Both TIM and TEM are driven by precession frequency of ions and electrons respectively,
and are generated in presence of ion and electron equilibrium gradients 
respectively \cite{drouot14:t, garbet:x2001}. However, in the 
nonlinear regime they depart from each other due to different response of
zonal flows \cite{ghizzo:a15,chen:a22, drouot15:t}. 
It is essential to properly estimate the linear growth rate 
of these trapped-particle instabilities to understand their influence on nonlinear saturation, 
the turbulent nature of the system, and the
associated transport. Although TEM is more directly relevant to tokamak 
turbulence and transport, TIM in numerical simulation is much more tractable and 
shares essential similarities with TEM in linear phase. Therefore here we 
focus on TIM instability.

During last few decades, ion turbulence in magnetically confined plasmas has been 
intensively studied in fluid simulations \cite{waltz94:r,ottaviani90:m,dorland:w}, 
fluid-kinetic hybrid electron model for studying low frequency electro-magnetic turbulence
\cite{lin:z01}, particle-in-cell based gyrokinetic simulations 
\cite{sydora:r,lee87:w,idomura:i}, 
continuum Vlasov approach gyrokinetic simulations in Eulerian grid \cite{manfredi96:g},
$\delta f$-particle-in-cell based simulations for bounce-averaged kinetic equations 
obtained by phase-space Lagrangian Lie-perturbation theory \cite{kwon:jae,fong:b},
and semi-Lagrangian based reduced bounce averaged gyro-kinetic simulations 
\cite{ghizzo:a,depret00:g,drouot14:t}. However, a kinetic model is necessary to 
capture the kinetic features of trapped ion modes,
and a full 5-dimension (5D) gyrokinetic model demands more computer resources, we 
consider a reduced 4-dimension (4D) gyro-kinetic model with averaging over 
gyro-motion and the banana-orbit motion, and adiabatic response of passing 
particles \cite{depret00:g}. This allows us to study the trapped ion mode (TIM) 
instability, driven through the resonant interactions of trapped ions with a wave, 
separately from the ITG instability which is generated in presence of ion 
pressure gradient. 

The trapped ion mode instability is driven by the ion temperature gradient 
$\nabla_{\perp}T_i$ and $\nabla B$ drift.
In the unfavorable curvature region of a tokamak, temperature gradient
is aligned with the magnetic field gradient, particles in the lower temperature 
region drift more slowly than in the higher temperature region, which yields a charge 
separation in the presence of a perturbed density profile. The electric field generated
from this charge separation creates a $E\times B$ drift motion, which enhances
the initial perturbation, hence leading to an instability. Though the nonlinear 
evolution of this TIM instability is interesting in the context of turbulent 
transport, it is first worth studying the linear properties of this instability.
A local-linear analysis of TIM instability using this reduced 
gyro-kinetic model was reported by Drouot {\it et. al.} \cite{drouot14:t}, 
where the values of all relevant parameters were considered at a particular 
radial location. The linear version of the simulation code TERESA 
\cite{depret00:g,sarazin:y,darmet08:g}, which is based on a semi-Lagrangian method, is
able to solve the modeled equations in the linear limit with radial profiles of all 
relevant parameters \cite{lesur:m}. Here we are presenting an
alternative {\it global-linear analysis} by solving the modeled equations of this 
reduced gyro-kinetic model in the linear limit using a spectral method 
\cite{gravier:e,gravier:e13}, which takes into account the radial profiles of 
all relevant parameters, and saving a lot of 
computational resources compared to the linear version of TERESA code. 

Moreover, till now, the trapped ion mode instability, and its growth-rate in the
presence of different temperature profiles and variations in precession frequency 
have been studied with the assumptions that the temperature profile varies linearly 
with $\psi$ and that the 
precession frequency remains constant throughout the simulation box, which is a good 
approximation for a simulation region, situated sufficiently far from the edge
( i.e. the last closed flux surface (LCFS) in our model) and the 
core region of a tokamak. Here using this solver, based on a spectral method,
we have studied the effects on the TIM instability of different temperature profiles $T_i(\psi)$, 
and precession frequency profiles $\Omega_D(\psi,\kappa)$ as a function of the magnetic 
flux surface $\psi$ and trapping parameter $\kappa$, which is associated with the 
{\it pitch angle} variation of trapped particles. All the temperature and precession 
frequency profiles are relevant to different experimental observations.
Depending on the temperature profiles and precession frequency values we found 
different growth rates of the TIM instability and different potential solutions of the system, 
which are significantly different
from the previous local linear analysis results \cite{drouot14:t}.  
Additionally for incorporating the variation of particles drift velocity with magnetic 
flux $\psi$, we have modified the previous 
gyro-bounce averaged kinetic model by introducing the $\psi$
dependency of the equilibrium Hamiltonian $H_{eq}$ in the quasi-neutrality equation 
and the equilibrium distribution function, which was neglected in the previous model 
where the equilibrium Hamiltonian was approximated by the energy $E$ of the particles: 
$H_{eq}\approx E$ \cite{sarazin:y}.
This new modification changes the growth-rate profile of TIM mode instability, and changes the
potential solution of the system. The 
growth-rate depends on the logarithmic gradients of temperature, density and 
equilibrium Hamiltonian, respectively $\kappa_{T}$, $\kappa_{n}$ and 
$\kappa_{\Lambda}$. All the results from this spectral method are successfully 
compared with the new linear version of TERESA, which
incorporates the $\psi$ and $\kappa$ dependency of precession frequency. 
Therefore this newly proposed method for global linear analysis of TIM 
instability, and the associated results help to 
understand the threshold/beginning of turbulent transport in tokamak by TIM modes, and 
also explain possible dominant mode of this instability in nonlinear regime.

Our paper is organized as follows, sec.~\ref{Bounce_avModel} presents the 
bounce-averaged gyrokinetic model, where a  brief description
of the previous nonlinear model, along with crucial upgrades are discussed.
Also the formulation of global linear analysis of this nonlinear 
model is presented in this section. Sec.~\ref{spectral_method} presents a numerical 
solver based on spectral method for solving the dispersion relation, 
which is derived from global linear 
analysis. The effect of different temperature profiles and precession frequency 
profiles on the TIM instability in the limit $H_{eq}\approx E$ is discussed in 
sec.~\ref{compare}. 
In sec.~\ref{LambdaD}, the effect of the inverse gradient length of equilibrium 
Hamiltonian $\kappa_{\Lambda}$, on the TIM instability is presented. 
Sec.~\ref{Conclusions} presents the conclusions. 
%
\section{The bounce-averaged gyrokinetic model and its modifications due to exact
Hamiltonian ($H(\psi,\kappa)$) \label{Bounce_avModel}}
An electrostatic reduced collisionless bounce-averaged gyrokinetic model was 
developed by Depret, Sarazin and Darmet 
\cite{depret00:g,drouot14:t,sarazin:y,darmet08:g}.
We adopt that model to study the stability of trapped ion modes.
In this model the system evolves on a timescale of the order of the trapped 
particle precession frequency $\omega_D$, which allows to filter out the 
large frequencies $\omega_c$ (cyclotron frequency) and $\omega_b$ 
(bounce frequency around banana orbit) ($\omega_D \ll \omega_b \ll \omega_c$), 
and to simplify the effect of length 
scales $\rho_c$ (gyro-radius) and $\delta_b$ (banana width). The dynamics 
of the gyro and bounce averaged trapped particles or {\it banana centre} 
distribution function $f_s$ is determined by the kinetic Vlasov equation,
\begin{eqnarray}
	\frac{\partial f_s}{\partial t} - 
	\left[ J_{0,s} \phi, f_s\right]_{\alpha,\psi}
	+\frac{\Omega_D(\psi,\kappa) E}{Z_s} \frac{\partial f_s}{\partial \alpha} = 0.
\label{vlasov}
\end{eqnarray}
where $\phi$ is the electrostatic potential. $Z_s$ is the charge number of the species. 
$[\cdots]_{\alpha,\psi}$ is the Poisson bracket in the phase space 
of toroidal precession angle $\alpha$ and poloidal magnetic flux $\psi$
($\psi \sim -r^2$, is used as a radial co-ordinate, where $r$ is radius of tokamak). 
The magnetic flux-function is calculated from the integration
$\psi(r)\sim -B_{min}\int_{0}^{r} \frac{r}{q_0}dr$, where the negative sign stems from
the convention that the direction of poloidal magnetic field $B_{\theta}$ and the 
normal direction of internal poloidal surface element is 
opposite to each other. Then
$\psi$ is shifted and normalized such that $\hat{\psi}$ is always positive, and 
$\hat{\psi}\in[0,1]$ corresponds to a limited radial extent of a tokamak with
$\hat{\psi}=0$ is associated with the edge, towards the last closed flux surface (LCFS),
and $\hat{\psi}=1$ is associated with the core, towards the hottest central region of a 
tokamak. This model is not valid inside the open field line regions,
where most of the trajectories intercept plasma-facing components. 
The normalization of $\psi$ and other essential physical quantities are given in 
Tab.~\ref{table:normaliz}, where dimensionless normalized quantities are noted with hat. 
However in the main text, the hat notation is omitted for clarity.
$E \Omega_D/Z_s = \omega_{d,s}$ is the energy dependent precession frequency of species $s$,
\begin{eqnarray}
	\omega_{d,s} = \frac{q(r)}{r}\frac{E}{q_s B_{min} R_0} \bar{\omega}_d, 
\label{omegaD}
\end{eqnarray}
where $q$, and $R_0$ are the safety factor, and major radius of 
a tokamak, respectively.
$q_s$ is the electric charge of the species $s$, $B_{min}$ is the minimal 
strength of the magnetic field on a field line, 
$E \equiv \frac{1}{2}m_s v_{G\parallel}^2 + \mu B_{G}$ is the particle kinetic 
energy, $\mu \equiv \frac{m_s v_{\perp}^2}{2B_G}$ is the magnetic moment, 
where subscript $G$ refers to the quantities computed at the position of 
guiding center, and
\begin{eqnarray}
\bar{\omega}_d = \frac{2\mathcal{E}(\kappa^2)}{\mathcal{K}(\kappa^2)}
	-1 + 4 s_0(r) \left(\frac{\mathcal{E}(\kappa^2)}{\mathcal{K}(\kappa^2)} 
+ \kappa^2 -1 \right),
\label{omegabar}
\end{eqnarray}
where $\kappa = \sqrt{\frac{1-\lambda}{2 \varepsilon \lambda}}$ is the 
trapping 
parameter which can vary between $0$ (for deeply trapped particles) to 
$1$ (at separatrix for barely trapped 
particle). $\varepsilon = a/R_0$ is the inverse of the aspect ratio of the 
tokamak, $a$ is the minor radius of tokamak, and for a tokamak with large aspect ratio 
usually the quantity $\varepsilon < 1$. 
$\lambda = \mu B_{min}(\psi)/E$ is the pitch angle, $s_0 =\frac{r}{q(r)}\frac{dq}{dr}$ 
is the magnetic shear, and $\mathcal{K}(\kappa^2)$ and $\mathcal{E}(\kappa^2)$ are 
the complete elliptic functions of the first and second kind, respectively. 
The operator $J_{0,s}$ performs two successive averages: the gyro-average and the 
bounce-average, which are the average over the cyclotron motion and the banana motion, 
respectively for the species $s$ is (according to ``Pad\'{e}'' expression
\cite{sarazin:y}). 
\begin{eqnarray}
\begin{aligned}
	J_{0,s} = \Bigg(\Bigg. 1 - \frac{E}{T_{eq,s}(0)}&\frac{\delta_{b0,s}^2}{4} 
	\partial^2_{\psi}\Bigg.\Bigg)^{-1} \\& \left( 1- \frac{E}{T_{eq,s}(0)}
	\frac{q^2 \rho_{c0,s}^2}{4 L_{\psi}^2} 
	\partial^2_{\alpha}\right)^{-1},
\end{aligned}
\label{gyro_bounce}
\end{eqnarray}
where $\rho_{c0,s} = \frac{m_s v_{\perp}}{q_sB}$ and 
$\delta_{b0,s}=q\rho_{c0,s}/\sqrt{\varepsilon}$ are the 
Larmor radius and the banana 
width (in unit of $\psi$) computed at temperature $T_0$. 
$T_{eq,s}(0)$ is the equilibrium temperature of species $s$ at $\psi=0$.

\begin{table}[h!]
\centering
\begin{tabular}{ c c c c c c c c c}
\hline
~Quantity~ &~ e.g.~ &~ Normalization~ \\
\hline
\hline
	~Time~ &~ $t$, $\omega^{-1}$~ &~ $\hat{t} = \omega_{d,0} t$~\\
	~Poloidal magnetic flux~ &~ $\psi$, $a$, $\rho_{c0}$, $\delta_{b}$~ &~ 
	$\hat{\psi}(r) = \frac{\psi(r)-\psi(a)}{L_{\psi}}$~\\
	~Electric potential~ &~ $\phi$ ~ &~ $\hat{\phi} = \phi/(\omega_{d,0}L_{\psi})$~\\
	~Energy~ &~ $E$ ~ &~ $\hat{E} = E/T_0$~\\
	~Density~ &~ $n_s$ ~ &~ $\hat{n}_s = n_s/n_0$~\\
	~Temperature~ &~ $T$ ~ &~ $\hat{T} = T/T_0$~\\
	~Distribution function~ &~ $f_s$, $F_{eq}$ ~ &~ $\hat{f}_s = 
	\frac{1}{n_0}\left(\frac{2\pi T_0}{m}\right)^{3/2} f_s$~\\
\hline
\hline
\end{tabular}
\caption{Normalization of the plasma parameters. Physical quantities are noted without 
	a hat, and
	dimensionless quantities are noted with a hat. Here $\omega_{d,0} = 
	q_{0}T_0/(e r_0 R_0 B_{0})$ is a typical precession frequency of strongly 
	trapped ion at $E=T_0$. $n_0$ and $T_0$ are arbitrary normalizing ion density
	and temperature such that $\hat{n}_s = \hat{T} = 1$ at $\hat{\psi} = 0$. The 
	quantity $L_{\psi}$ is the radial size of the simulation box in magnetic flux 
	unit. The minor radius $a$, the Larmor radius $\rho_{c0}$, and the 
	banana width $\delta_b$ are all expressed in units of $\psi$. However in the 
	main text, the hat notation is omitted for clarity.}
\label{table:normaliz}
\end{table}

Self-consistency is ensured by a quasi neutrality constraint, including a 
polarization term $\bar{\Delta}_s\phi$, where $\bar{\Delta}$ is a non-isotropic
Laplacian operator,
\begin{eqnarray}
	\bar{\Delta}_s = \left(\frac{q \rho_{c0,s}}{L_{\psi}} \right)^2
	\frac{\partial^2}{\partial \alpha^2} + \delta^2_{b,s} 
	\frac{\partial^2}{\partial \psi^2}.
\label{Laplacian}
\end{eqnarray}
where $L_{\psi} = a R_0 B_{\theta}$ is the radial length of the simulation box 
in unit of $\psi$. 
The quasi-neutrality equation with the approximation 
$H_{eq,s}(\psi) = E(1+\Omega_{D}\psi) \approx E$
reads as follows \cite{sarazin:y,drouot:t},
\begin{eqnarray} 
\begin{aligned}
	\frac{2}{\sqrt{\pi} n_{eq}(0)}&\sum_{s} Z_s 
	\int_0^{1}\kappa\mathcal{K}(\kappa^2) d\kappa
	\int_{0}^{\infty}
	J_{0,s} f_s \sqrt{E} dE 
	\\&= \sum_{s} \frac{e Z_s^2}{T_{eq,s}(0)}
	\left[ \frac{1-f_t}{f_t}(\phi - \epsilon_{\phi,s}
	\langle\phi\rangle_{\alpha}) -\bar{\Delta}_s \phi\right],
\end{aligned}
\label{quasiNutrality}
\end{eqnarray}
where, $n_{eq}(0)$ is the equilibrium density at $\psi=0$, 
$f_t = \frac{2\sqrt{2\varepsilon}}{\pi}$ is the fraction of trapped 
particles, which scales as $\sqrt{r}$, whereas we consider a constant $f_t$.
$\langle\hdots\rangle_{\alpha}$ corresponds to the average on the angle $\alpha$.
Here passing particles are treated quasi-adiabatically.
In all the previous studies of trapped particle modes instability 
\cite{depret00:g,sarazin:y,darmet08:g},
the gyro-bounce averaged kinetic model was simplified by considering the equilibrium 
Hamiltonian 
$H_{eq}(\psi) = E (1+\Omega_D \psi) \approx E $, which was used to derive the 
quasi-neutrality equation 
Eq.~(\ref{quasiNutrality}). 
This approximation is valid only at 
$\psi =0$ and/or $\Omega_D\ll 1$. The exact expression for equilibrium Hamiltonian is, 
\begin{eqnarray}
	H_{eq}(\psi,\kappa) = E\left( 1+\int_0^{\psi}\Omega_D(\tilde{\psi},\kappa) 
	d\tilde{\psi}\right) 
	= E \Lambda_D(\psi,\kappa), 
\label{Heq_Lambda}
\end{eqnarray}
where $\Lambda_D(\psi,\kappa) = 1+ \int \Omega_D(\tilde{\psi},\kappa) d\tilde{\psi}$, 
with $\tilde{\psi}$ is a variable of integration. 
In the previous derivation of gyro-bounce averaged Vlasov equation 
Eq.~(\ref{vlasov}) the $\psi$ derivative of the equilibrium Hamiltonian 
was considered as $\frac{dH_{eq}}{d\psi} = E\Omega_D$, which remains unchanged 
for this new Hamiltonian Eq.~(\ref{Heq_Lambda}). Therefore the 
expression for gyro-bounce averaged Vlasov equation Eq.~(\ref{vlasov}) remains 
unchanged. With this new equilibrium Hamiltonian Eq.~(\ref{Heq_Lambda})
the elementary volume in the phase-space can be 
written as $d^3v = 4\pi \sqrt{2} m^{-3/2} \sqrt{E} \Lambda_D^{3/2} dE 
\frac{d\lambda}{4\Omega_D}$. 
For simplicity we consider, the term $\bar{\Delta}\phi$ associated with the 
polarization in the quasi-neutrality equation Eq.~(\ref{quasiNutrality}) remain 
unchanged under this new modification in the equilibrium Hamiltonian. 
By keeping unchanged the right-hand side of the previous quasi-neutrality equation
Eq.~(\ref{quasiNutrality}) and using the modified expression of elementary volume 
$d^3v$ for the integration of gyro-bounce averaged distribution $f_s$, 
the quasi-neutrality equation can be written as 
\begin{eqnarray} 
\begin{aligned}
	\frac{2\Lambda_{D}(\psi,\kappa)^{3/2}}{\sqrt{\pi} n_{eq}(0)}\sum_{s} Z_s 
        \int_0^{1}\kappa\mathcal{K}(\kappa^2) d\kappa
        \int_{0}^{\infty}
        J_{0,s} f_s \sqrt{E} dE 
	\\ = \sum_{s} \frac{e Z_s^2}{T_{eq,s}(0)}
        \left[ \frac{1-f_t}{f_t}(\phi - \epsilon_{\phi,s}
        \langle\phi\rangle_{\alpha}) -\bar{\Delta}_s \phi\right],
\end{aligned}
\label{quasiNutrality_mod}
\end{eqnarray}
Therefore there is an additional multiplication term $\Lambda_D(\psi,\kappa)^{3/2}$ 
that arises in the left-hand side of the previous quasineutrality condition 
Eq.~(\ref{quasiNutrality}). In this model, the gyro-bounce averaged distribution 
function $f_s$ bears $4$ dimensions ($\alpha,\psi,E,\kappa$), and in a further 
reduced limit of a single $\kappa$ value it reduces to a $3-$dimension model.
%
\subsubsection{Global linearised model \label{Global_linear}}
%
Considering an initial very small amplitude perturbation of gyro-bounce 
averaged distribution $f_s = F_{eq,s} + \tilde{f}_s$ and potential 
$\phi = \tilde{\phi}$ in Fourier space as 
$\tilde{f}_s = \sum_{n,\omega} f_{s,n,\omega}(\psi,E,\kappa) \exp\{i(n\alpha -\omega t)\}$
and $\tilde{\phi} = \sum_{n,\omega} \phi_{n,\omega}(\psi) 
\exp\{i(n\alpha -\omega t)\}$, then neglecting the higher order nonlinear terms
$\left[ J_{0,s} \tilde{\phi}, \tilde{f}_{s}\right]_{\alpha,\psi}$,
the linearised form of Vlasov equation can be written as,
\begin{eqnarray}
	\frac{\partial \tilde{f}_s}{\partial t} - 
	\left[ J_{0,s} \tilde{\phi}, F_{eq,s}\right]_{\alpha,\psi}
     +\frac{\Omega_D E}{Z_s} \frac{\partial \tilde{f}_s}{\partial \alpha} = 0.
\label{linear-vlasov}
\end{eqnarray}
$n$ and $\omega$ are the mode number (along $\alpha$), and angular frequency of the 
Fourier modes respectively.
Here we consider the normalized gyro-bounce averaged equilibrium distribution 
$F_{eq,s}$ is Maxwellian energy distribution, which is independent of 
($\alpha, t$). 
Using the new equilibrium Hamiltonian $F_{eq,s}$ can be written as:
\begin{eqnarray}
        F_{eq,s}(\psi,E,\kappa) = \frac{n_{eq,s}(\psi)}{T_{eq,s}^{3/2}(\psi)} 
	\exp\left(-\frac{E \Lambda_D(\psi,\kappa)}{T_{eq,s}(\psi)}\right), 
\label{Fequilibrium_Lambda}
\end{eqnarray}
where $T_{eq,s}(\psi)$ and $n_{eq,s}(\psi)$ are the temperature and density profiles 
of the equilibrium distribution function of species $s$, respectively.
In this case, the integration of $F_{eq}$ over the velocity space $d^3v$ leads 
to $n_{eq}(\psi)$. Hereafter we denote $T_{eq}(\psi)$ as $T(\psi)$ and 
$n_{eq,s}(\psi)$ as $n_s(\psi)$.
The term $E\Lambda_D$ allow us to incorporate the radial
variation of particle drift velocity.
After substituting $\tilde{f}_{s}$, $\tilde{\phi}$ and $F_{eq,s}$ in 
Eq.~(\ref{linear-vlasov}), the solution of Vlasov equation in Fourier space become,
\begin{eqnarray}
\begin{aligned}
        f_{n,\omega} =& 
	\frac{ \kappa_n(\psi) + \kappa_{T}(\psi)
	\left(\frac{E \Lambda_D(\psi)}{T(\psi)}-\frac{3}{2}\right)
	-\frac{E\Lambda_D}{T(\psi)}\kappa_{\Lambda}(\psi)}
	{Z_s^{-1}\Omega_{D}(\psi,\kappa) E-\frac{\omega}{n}} \\& 
        \left\{J_{0,n,s}\phi_{n,\omega}(\psi)\right\} F_{eq,s}(\psi,E,\kappa).
\end{aligned}
\label{linearfSol_Lambda}
\end{eqnarray}
where $\kappa_n(\psi) = \frac{1}{n_s(\psi)}\frac{dn_s}{d\psi}$,
$\kappa_T(\psi) = \frac{1}{T_s(\psi)}\frac{dT_s}{d\psi}$ and 
$\kappa_{\Lambda}(\psi) = \frac{1}{\Lambda_D(\psi)}\frac{d\Lambda_D}{d\psi} 
=\frac{\Omega_{D}(\psi)}{\Lambda_D(\psi)}$
are the logarithmic gradients of density, temperature and 
equilibrium Hamiltonian $H_{eq}(\psi)$, respectively.

Considering both electron and ion contributions,
the quasi-neutrality condition Eq.~(\ref{quasiNutrality_mod}) can be written as,
\begin{eqnarray} 
\begin{aligned}
	&\frac{\sqrt{\pi}}{2 T_{i}(0)} \left[C_{ad}(\phi_{n,\omega} -
	\epsilon_{\phi}\langle\phi\rangle_{\alpha}) 
	- C_{pol}\bar{\Delta}\phi_{n,\omega}\right]  
	= \mathcal{N}_{n,i} -\mathcal{N}_{n,e}, \\&
	\mathcal{N}_{n,i} = \frac{\Lambda_{D}^{3/2}}{n_i(0)}	\int_0^{1}
	\int_{0}^{\infty}J_{0,i}f_{i,n,\omega}(\psi,E,\kappa)\sqrt{E} dE
	\kappa\mathcal{K}(\kappa^2) d\kappa
	, \\ &
	\mathcal{N}_{n,e} = \tau \frac{\Lambda_{D}^{3/2}}{n_e(0)}\int_0^{1}
	\int_{0}^{\infty}J_{0,e}f_{e,n,\omega}(\psi,E,\kappa)\sqrt{E} dE 
	\kappa\mathcal{K}(\kappa^2) d\kappa,
\end{aligned}
\label{quasi-Nutr2}
\end{eqnarray}
where $\tau = \frac{T_{i}}{T_{e}}\big|_{\psi=0}$, $C_{pol} = \frac{q_i\omega_{0}L_{\psi}}{T_0}$, 
$C_{ad} = \frac{1-f_t}{f_t}(1-\tau) C_{pol}$, 
$\epsilon_{\phi} = \frac{\epsilon_{\phi,i}+\tau\epsilon_{\phi,e}}{1+\tau}$ and
$\bar{\Delta}\phi = \bar{\Delta}_{i}\phi + \tau \bar{\Delta}_{e}\phi$. 
In the limit of a constant {\it pitch-angle},
the value of $\Omega_D$ is constant along $\kappa$, then
the $\kappa$ integration in Eq.~(\ref{quasi-Nutr2}) can be simplified as, 
$\int_0^1\kappa\mathcal{K}(\kappa^2) d\kappa = 1$.
Substituting the value of $f_{n,\omega}$ from Eq.~(\ref{linearfSol_Lambda}) in 
Eq.~(\ref{quasi-Nutr2}), the expression of $\mathcal{N}_s$ becomes 
\begin{eqnarray}
\begin{aligned}
	&\mathcal{N}_{n} = \frac{\Lambda_D(\psi)^{3/2}}{n_s(0)}\int_{0}^{\infty}
        \sqrt{E} J_{0,n}\Bigg[\Bigg.
        \frac{n_s(\psi)}{T^{3/2}(\psi)} 
	\exp\left(-\frac{E\Lambda_D}{T(\psi)}\right)
	\\&
	\frac{ \kappa_{n}(\psi) + \kappa_{T}(\psi)
	\left(\frac{E \Lambda_D}{T(\psi)}-\frac{3}{2}\right)
	-\kappa_{\Lambda}(\psi)\frac{E\Lambda_D}{T(\psi)}}
	{Z^{-1}\Omega_{D}(\psi) (E-\chi)} 
	(J_{0,n}\phi_{n,\omega})
        \Bigg.\Bigg] dE, 
\end{aligned}
\label{Ns_res_LambdaD}
\end{eqnarray}
where $\chi = \frac{\omega}{n Z^{-1}\Omega_D(\psi)}$, 
and $\omega$ has both real and imaginary parts $\omega = \omega_r + i\gamma$. 
Due to the term
$(E-\chi_s)$ in denominator there is a possibility of resonance between wave 
and particle motion. For ions $Z_i$ is positive, therefore the resonance occurs
only when the phase velocity of wave has the same sign as the ion precession drift
(ie., $\omega > 0$). By substituting 
$\bar{\Delta}_s$ from Eq.~(\ref{Laplacian}) in the left-hand side of 
quasineutrality condition Eq.~(\ref{quasi-Nutr2}) we define the differential operator 
\begin{eqnarray}
\begin{aligned}
	C_n =& \frac{\sqrt{\pi}}{2 T_{i}(0)}\Bigg[\Bigg. C_{ad}
	(1+\epsilon_{\phi}\delta_{n,0})\\&
	+C_{pol}\left\{({\rho^{*}_{i}}^2 +\tau{\rho^{*}_{e}}^2)
	(-n^2) - (\delta_{bi}^2+\tau\delta_{be}^2)
	\frac{\partial^2}{\partial\psi^2} \right\}\Bigg.\Bigg],
\end{aligned}
\label{CnDefine}
\end{eqnarray}
where $\rho^{*}_{s} = \frac{q\rho_{c0,s}}{L_{\psi}}$, and $\delta_{n,0}$ is Kronecker 
delta with the value $1$ for $n=0$ and for $n\ne 0$, it is $0$. 
Therefore the dispersion relation becomes,
\begin{eqnarray}
	C_n\phi_{n,\omega} = \mathcal{N}^*_{n,i} \phi_{n,\omega} 
	-\mathcal{N}^*_{n,e} \phi_{n,\omega} 
\label{Dispersion}
\end{eqnarray}
where $\mathcal{N}^*_{n,s} = \frac{\mathcal{N}_{n,s}}{\phi_{n,\omega}}$ is actually
a differential operator (Eq.~(\ref{gyro_bounce}) and (\ref{Ns_res_LambdaD})) acting on 
$\phi_{n,\omega}$. We will come back to this issue 
in Sec.~\ref{spectral_method}. Hereafter we remove the subscript $\omega$ from 
$\phi_{n,\omega}$, and denote it as $\phi_{n}$, because the linear dispersion 
relates $\omega$ to $n$ values. By substituting the expression of $C_n$ in 
Eq.~(\ref{Dispersion}) one can derive a 
2nd order linear differential equation of $\phi_{n,\omega}$ as,
\begin{eqnarray}
\begin{aligned}
	\frac{d^2\phi_n}{d \psi^2} &+ Q_n(\psi)\phi_{n} = 0, \\
	Q_n(\psi) &= \frac{\mathcal{N}^*_{n,i} -\mathcal{N}^*_{n,e} 
	-\frac{\sqrt{\pi}}{2 T_{i}(0)}\left[C_{ad}+C_{pol}({\rho^{*}_{i}}^2 
	+\tau{\rho^{*}_{e}}^2)n^2 \right]}{\frac{\sqrt{\pi}}{2 T_{i}(0)} 
	C_{pol}({\delta_{bi}}^2 +\tau{\delta_{be}}^2)}.
\end{aligned}
\label{QMatrix}
\end{eqnarray}
Considering $T_{i}(0) = T_0$, $n_i(0) = n_e(0) = n_0$, in normalized unit those become
$T_{i}(0) = n_s(0) = 1$.
Here we study the modes for which $n \ne 0$, therefore the term 
$\epsilon_{\phi}\delta_{n,0}$ in Eq.~(\ref{CnDefine}) vanishes. 
For simplicity to study specifically the trapped ion mode instability (TIM) 
we will neglect the electron perturbation $\tilde{f}_e = 0$ which leads to 
$\mathcal{N}^*_e =0$. However, the same method is applicable to trapped electron mode 
(TEM) instability by setting $\mathcal{N}^*_i =0$ and $\mathcal{N}^*_e\neq 0$. 
Moreover, for $\tau \le 1$, 
${\rho_i^*}^2\gg {\rho_e^*}^2$ and $\delta_{bi}^2\gg \delta_{be}^2$. 
Therefore after neglecting those terms 
associated with electrons contribution and substituting $T_{eq,i}=1$ 
in eq.~(\ref{QMatrix}), the modified expression for $Q_n(\psi)$ becomes,
\begin{eqnarray}
        Q_n(\psi) = \frac{\mathcal{N}^*_{n,i}  
	-\frac{\sqrt{\pi}}{2}\left[C_{ad}+C_{pol}{\rho^{*}_{i}}^2 
	n^2 \right]}{\frac{\sqrt{\pi}}{2} C_{pol}
        {\delta_{bi}}^2 }
\label{QMatrix1sp}
\end{eqnarray}
The contributions of the density gradient, temperature gradient and the 
gradient in the equilibrium Hamiltonian is contained in the expression of 
$\mathcal{N}^*_n$. Due to the term $-\kappa_{\Lambda}$, while the increase in 
logarithmic gradients of temperature and density $\kappa_T$ and $\kappa_n$ 
help to enhance the TIM instability, the logarithmic gradient of equilibrium
Hamiltonian $\kappa_{\Lambda}$ helps to stabilize the TIM instability.
In the limit $H_{eq}\approx E$, the term $\kappa_{\Lambda}= 0$ and 
$\Lambda_D=1$, in Eq.~(\ref{Ns_res_LambdaD}).
Therefore the growth-rate $\gamma$ of all the modes $n$ of TIM instability 
for the new modified equilibrium Hamiltonian $H_{eq}(\psi,\kappa,E)$, will be
significantly smaller compared to the case with the limit $H_{eq}\approx E$.  
The dispersion relation of the TIM mode instability in the limit of 
$H_{eq}\approx E$ is presented in the Appendix~\ref{Appd_HeqE}.
Since the solution in Fourier space Eq.~(\ref{linearfSol_Lambda}) 
$f_{n,\omega} = 0$ for $n=0$, the linear 
analysis is valid only for the mode numbers $n > 0$. However, $n=0$ mode is 
linearly stable, it cannot extract free energy from the equilibrium gradients.  
%
\subsubsection{Local linearised analysis \label{local_linear}}
%
The local linear stability analysis of this reduced gyro-bounce 
averaged model with the new modified equilibrium Hamiltonian can be obtained by
expanding $F_{eq,s}(\psi,\kappa,E)$ eq.~(\ref{Fequilibrium_Lambda}) up to 
$1^{\rm st}$ order of 
$\psi$ around $\psi= 0$, and substituting $\partial_{\psi}^2 \phi = -k^2 \phi$ 
(with $k = \pi$ ) 
in the dispersion relation, which leads to the new simplified dispersion relation as,
$C_n-\mathcal{N}^*_{n,i} = 0$.
From this, using Plemelj formula \cite{plemelj:j}, the threshold frequency 
value of the real part of frequency $\omega_r$ for TIM 
instability can be derived as:
\begin{eqnarray}
	\omega_r^{new} = \left(\frac{\frac{3}{2} \kappa_{T0} -\kappa_{n0}}
	{\kappa_{T0} - 
	\kappa_{\Lambda 0}}\right) \frac{\Omega_{D0}}{\Lambda_D} T_0 n,
\label{Omegar_loc_LambdaD}
\end{eqnarray}
where the subscript `$0$' denotes the variables value at $\psi=0$. 
Here we restrict the analysis to the case of resonant interactions only, i.e.
$\omega_r > 0$, hence Eq.~(\ref{Omegar_loc_LambdaD}) is valid for the cases
with $(3 \kappa_T/2-\kappa_n) / (\kappa_T-\kappa_{\Lambda}) >0$. 
The threshold value of $\kappa_T$ for the instability can be written as
\begin{eqnarray}
	\kappa_{T,th}^{new} = \frac{C_n\Omega_{D0}}{\Lambda_{D0} \int_0^{\infty} J_{0,n}^2 
	\exp(-\xi)\sqrt{\xi} d\xi} +\kappa_{\Lambda 0}
\label{KT_thr_LambdaD}
\end{eqnarray}
where $\xi = \frac{E\Lambda_D}{T}$. The threshold values of $\omega_r$ and
$\kappa_T$ Eq.~(\ref{Loc_linHE}-\ref{KT_thrHE}) in the limit of $H_{eq}\approx E$ 
\cite{drouot14:t} can be recovered by substituting 
$\Lambda_D =1$ and $\kappa_{\Lambda}=0$ in 
Eq.~(\ref{Omegar_loc_LambdaD}-\ref{KT_thr_LambdaD}). 
The relation between these two threshold $\kappa_T$ values is
\begin{eqnarray}
	\kappa_{T,th}^{new} = \frac{\kappa_{T,th}}{\Lambda_{D0}}+ 
	\kappa_{\Lambda 0},
\label{KT_thr_rel_LambdaD}
\end{eqnarray}
where $\kappa_{T,th}^{new}$ and $\kappa_{T,th}$ are the threshold values of 
$\kappa_T$ for TIM instability in the case with $\kappa_{\Lambda}$ and without 
$\kappa_{\Lambda}$, respectively. For $\Lambda_{D0}\approx 1$, $\kappa_{T,th}^{new} \gg 
\kappa_{T,th}$, which suggests a relatively strong gradient in temperature 
profile is required in order to obtain TIM instability, compared to the case with
$H_{eq}\approx E$. In case of adiabatic electron response 
there is a non-resonant branch of TIM for $\chi < 0$ \cite{sarazin:y}. But here we are 
focusing only on the resonant branch of TIM, for which $\chi > 0$.

In the present study, unlike the local linear analysis, instead of Taylor expansion of 
$F_{eq,s}$ around 
$\psi=0$, a full $\psi$ dependent Maxwellian distribution for the equilibrium 
gyro-bounce averaged particle distribution $F_{eq,a}(\psi,\kappa,E)$ is considered.
Therefore using this global linear analysis, effects of any type of 
temperature, density and precession frequency profiles on the stability of trapped 
ion modes (TIM) can be investigated.
In the next section we will solve the differential equation of $\phi$ 
eq.~(\ref{QMatrix}) with the expression for $Q_n$ from 
Eq.~(\ref{QMatrix1sp}) using spectral method. 
%
\section{Global linear analysis: Spectral method \label{spectral_method}}
%
The $Q_n$ function in Eq.~(\ref{QMatrix1sp}) depends on $\psi$ in a intricate 
manner. However in the limit of local linear analysis \cite{drouot14:t}, 
$Q_n$ is independent of $\psi$. Therefore $-Q_n$ becomes the eigenvalue of 
the operator $\partial_{\psi}^2$ in Eq.~(\ref{QMatrix}). 
But in our global linear analysis, due to full Maxwellian distribution 
$F_{eq}$, it is impossible to eliminate the $\psi$ dependency in $Q_n$, 
ie., $-Q_n$ is no more an eigenvalue of $\partial_{\psi}^2$. Here we use 
a spectral approach \cite{gravier:e,gravier:e13,pryimak:v} to solve 
Eq.~(\ref{QMatrix}). We can consider our co-ordinate system $(\psi,\alpha)$ 
as a polar co-ordinate system with
$\psi$ as radial axis and $\alpha$ as angular axis. If we consider the solution 
$\phi(\psi,\alpha)$ as a scalar analytic function in the region $0\leq\psi\leq 1$, 
according to the Theorem-1 of the page-$374$ in \cite{pryimak:v}, using a spectral 
approach the solution $\phi(\psi,\alpha)$ can be expanded as:
\begin{eqnarray}
	\phi(\psi,\alpha) =\sum_{n=-\infty}^{\infty}\psi^{|n|}\sum_{m=1}^{\infty} 
	\hat{\phi}_{n,m} \mathfrak{F}_{m}(\psi)\exp(i n\alpha)
\label{spectotal}
\end{eqnarray}
where $\mathfrak{F}_{m}(\psi)$ is an even power function of $\psi$, and the sum index 
$m$ has values $m= 1,2\cdots\infty$. In practice, a truncation is performed so 
that it is limited to the first $M$ number functions of $\mathfrak{F}_m$. 
In Eq.~(\ref{QMatrix}) all the differentiation in $\alpha$ is taken in Fourier-space. 
Therefore according to the 
spectral approach from Eq.~(\ref{spectotal}) the solution of Eq.~(\ref{QMatrix}) 
$\phi_n(\psi)$ can be expressed as:
\begin{eqnarray}
        \phi_n(\psi) =\psi^{|n|}\sum_{m=1}^{M}
        \hat{\phi}_{n,m} \mathfrak{F}_{m}(\psi)
\label{phiexpand}
\end{eqnarray}
The boundary conditions for $\phi_n(\psi)$ in Eq.~(\ref{QMatrix}) are 
$\phi|_{\psi=0} = 0$ and $\phi|_{\psi = 1} = 0$. 
Our first choice for $\mathfrak{F}_{m}$ was
$\mathfrak{F}_{m} = (1-\psi^2)\psi^{2(m-1)}$, but this choice gives poorly 
conditioned matrices. For good conditions matrices the coefficients 
$\hat{\phi}_{n,m}$ go down towards zero exponentially with increasing $m$ and 
at $m=M$, $\hat{\phi}_{n,M}\approx 0$.
Then we consider orthogonal polynomials, Chebyshev polynomials of first kind 
$T_{2(m-1)}(\psi)$, for constructing $\mathfrak{F}_{m}$ as:
\begin{eqnarray}
	\mathfrak{F}_{m}(\psi) = (1-\psi^2)T_{2(m-1)}(\psi).
\label{fmn}
\end{eqnarray}
which were used for studying the collisional drift wave and 
ITG instabilities in a cylindrical plasma, using spectral method 
\cite{gravier:e,gravier:e13}. However, in 
case of TIM instability we found that this choice of function gives good 
conditions matrices for lower values of $n$. For higher values of $n$, the 
coefficients $\hat{\phi}_{n,m}$ 
do not go down towards zero exponentially with increasing $m$.   
Then we choose $n=1$ for the factor $\psi^{|n|}$, in the expansion function 
Eq.~(\ref{phiexpand}) and the dependency of solution on the mode-number $n$ is 
entirely determined by the expression $Q_n$ in Eq.~(\ref{QMatrix}). 
With this new choice the coefficients $\hat{\phi}_{n,m}$ go down towards zero 
exponentially with increasing $m$. Moreover we have checked that, using this 
new choice of expansion function, one can recover the solutions of 
collisional drift wave instability study, as given in the references 
\cite{gravier:e,gravier:e13}.   

By defining $\mathcal{C}_m = \psi\mathfrak{F}_{m}$ 
the expression of $\phi_n(\psi)$ can be written as,
\begin{eqnarray}
	\phi_n(\psi) = \sum_{m=1}^{M}\hat{\phi}_{n,m}\mathcal{C}_m.
\label{fmConstract}
\end{eqnarray}
Here $\mathcal{C}_m = \psi(1-\psi^2)T_{2(m-1)}(\psi)$. Using this new choice of 
$\phi_n(\psi)$  rather good conditioning properties of matrices are obtained for large
values of $n$ in our study. 
Substituting the value of $\phi$ in Eq.~(\ref{QMatrix}) we get,
\begin{eqnarray}
	\sum_{m=1}^M \hat{\phi}_{n,m} \frac{d^2\mathcal{C}_m(\psi)}{d\psi^2} =
	-\sum_{m=1}^M \hat{\phi}_{n,m} Q_n(\psi)\mathcal{C}_m(\psi),
\label{Q_spetral}
\end{eqnarray}
where $m$ is the index of the spectral function $\mathcal{C}_m$, and $n$ is wave 
number of TIM instability along the toroidal precession angle $\alpha$. 
The $2^{nd}$-order differential Eq.~(\ref{Q_spetral}), is solved numerically.
By proceeding this way, Eq.~(\ref{Q_spetral}) is only solved at the points $\psi$, 
where its left hand side is vanishing for the first dropped higher order term of 
the expansion, namely the term $\mathcal{C}_{M+1}(\psi) = \psi(1-\psi^2)T_{2M}(\psi)$. 
Those $\psi$ values are called the collocation points, which are defined by the $M$ 
zeros of the Chebyshev polynomial $T_{2M}(\psi)$ situated in the interval
$0\leq\psi\leq 1$, ie., $\psi_l = \cos\left(\frac{(2l-1)\pi}{4M}\right)$
for $l\in \{1,\cdots,M\}$. Since $\mathcal{C}_{M+1}$ vanishes at these $\psi$ 
locations, for $J_0 =1$ the right hand side of Eq.~(\ref{Q_spetral}) 
also vanishes at $m=M+1$. Moreover, for gyro-bounce average operator $J_0$ according to 
Eq.~(\ref{gyro_bounce}) we solve the Eq.~(\ref{Q_spetral}) iteratively where the 
operator $\mathcal{N}^{*}_n$ inside $Q_n(\psi)$ is applied on a solution 
$\phi_{n-\Delta n}(\psi)$ which is obtained from the previous iteration. Therefore, at any 
iteration $Q_n(\psi)$ is a known function of $\psi$, it does not directly operate 
on $\mathcal{C}_m$. Hence at the collocation points the 
right hand side of Eq.~(\ref{Q_spetral}) vanishes for the $m=(M+1)$. 
For minimizing the truncation error, the discretization in $\psi$ is 
considered at those collocation points. We will explain this iteration method at 
the end of this section. For each value of wave number $n$ along $\alpha$, 
Eq.~(\ref{Q_spetral}) is evaluated at these collocation points and yields the 
matrix problem, 
\begin{eqnarray}
	\mathcal{M}_{D} \hat{\phi}_n = -\mathcal{M}_{Q}\hat{\phi}_n
\label{MEq}
\end{eqnarray}
where $\hat{\phi}_n = \left( \begin{array}{c} \hat{\phi}_{n,1} \\\vdots\\
\hat{\phi}_{n,M} \end{array} \right)$
is the vector representing the solution, and the matrices are,
\begin{eqnarray}
	(\mathcal{M}_{D})_{l,m} = \frac{d^2}{d\psi_l^2}\mathcal{C}_{m}(\psi_l),\\
	(\mathcal{M}_{Q})_{l,m} = Q_n(\psi_l)\mathcal{C}_{m}(\psi_l),
\label{Matx}
\end{eqnarray}
with $l,m \in \{1,\cdots,M\}$. Then we scan the $(\omega_r,\gamma)$ plane (where
$\omega= \omega_r +i\gamma)$,
and search for values of $\omega$ such that one eigenvalue of the matrix
$\mathcal{M} = \mathcal{M}_D + \mathcal{M}_Q$ vanishes within machine 
precision. During this search, there is no restriction on the range of 
$\omega_r$ values, and we search only for positive $\gamma$ values. 
For this purpose a method that finds the minimum of a scalar 
function of several variables, starting at an initial guess value, and 
iterates using the {\it simplex search} method \cite{lagarias:j} is used. 
Then using the eigenvector of matrix $\mathcal{M}$ associated with this smallest 
eigenvalue, the electric potential $\phi_n(\psi)$ can be constructed using 
Eq.~(\ref{fmConstract}). The solution $\phi_n(\psi)$ has both real and imaginary part.
The {\it spectral-convergence} or rate
of change of the coefficients $\hat{\phi}_{n,m}$ values with different $M$ values 
depends on the nature of the matrix $\mathcal{M}_Q$. 
Usually acceptable smooth $\hat{\phi}_{n,m}$s follows 
exponentially decaying functions with $m$, i.e., 
$\hat{\phi}_{n,m} = \hat{\phi}_{n,0}\exp(-|\beta| m)$, where $\beta$ is a  
constant. For different initial guess values the minimum 
searching method can converge towards different values of $\omega$.
Among these solutions one finds the couple ($\omega_r,\gamma$) for which 
the instability growth rate $\omega_i$ is maximum. We will take that 
particular set as a solution ($\omega_r, \gamma$).   
For generating the curve ($\omega_r$ Vs. $n$)  and 
($\gamma$ Vs. $n$) we start from mode $n=1$ and search for the highest value of 
growth rate, by choosing different initial guess values as 
$\omega_{r0}\in [0,3 \omega_{rL}]$ and $\gamma_{0}\in [0,\omega_{rL}]$ (since 
$\omega_r\ge\gamma$), and with the intervals 
$\Delta \omega_{r0}= 0.1 \omega_{rL}, \Delta\gamma_0 = 0.01 \omega_{rL}$,
where $\omega_{rL}$ is the threshold value of $\omega_r$ obtained from the local 
linear analysis Eq.~(\ref{Omegar_loc_LambdaD}).
Actually the range and the intervals for the initial guess values depend on 
the nature of instability ie., on the matrix $\mathcal{M}_Q$. After finding 
the highest value of 
growth rate for $n=1$, increase the $n$ value slowly with a step $\Delta n = 0.1$. 
Though the mode numbers $n$ along $\alpha$, are integer number, there is no such 
restriction on $n$ for solving the differential Eq.~(\ref{Q_spetral}) for fractional 
values of $n$.
In the search method, we choose the solution $\omega$ of $n-\Delta n$ step, as initial guess 
($\omega_{r0}, \gamma_{0}$) value to search the solution for mode number $n$. Here 
we consider the ($\omega_r$ Vs. $n$) and 
($\gamma$ Vs. $n$) profiles vary smoothly with $n$, such that small change in $n$, 
changes the $\mathcal{M}$ matrix by small amount, and helps to find-out the 
values $(\omega_r,\gamma)$ close to the initial guess value $(\omega_{r0},\gamma_{0})$,
 and keep the solutions within the same branch with highest growth rate.
Larger step in $\Delta n$ in the searching method, may lead to a departure of the 
solution ($\omega_r, \omega_i$) from the branch with highest growth rate. 
However if the solution departs from any particular branch, i.e. if it
leads to a sudden change in potential profile for 
two consecutive $n$ values, then 
one has to decrease the $\Delta n$ value. 

The main difficulty for solving the Eq.~(\ref{QMatrix}) using the 
spectral method is to construct the $Q_n(\psi)$ matrix. Because for 
calculating $\mathcal{N}^*_{n,i}$, the gyro-bounce average operator 
$J_{0,n,i}$ is applied to the potential solution $\phi_n$ in 
Eq.~(\ref{Ns_res_LambdaD}), which is an unknown quantity of Eq.~(\ref{QMatrix}). 
Therefore we will solve the Eq.~(\ref{QMatrix}) for $\phi$ iteratively.
Moreover to calculate 
the gyro-bounce average of any function $\mathcal{F}(\psi)$ using the Pad\'{e} 
expression Eq.~(\ref{gyro_bounce}), we solve the differential equation,
\begin{eqnarray}
	\left( 1-\frac{E}{T(0)}\frac{\delta_{b0,s}^2}{4} 
	\frac{d^2}{d \psi^2}\right)
	\Bar{\Bar{\mathcal{F}}}(\psi) = 
	\left(1+\frac{E}{T(0)}\frac{{\rho*}^2}{4} n^2\right)^{-1}
	\mathcal{F}(\psi),
\label{Gyro_averageSpectral}
\end{eqnarray}
where, $\mathcal{F}$ is a known function, on which the gyro-bounce average 
operator $J_{0}$ is applied and generates the gyro-bounce averaged quantity 
$\bar{\bar{\mathcal{F}}}$. We solve the Eq.~(\ref{Gyro_averageSpectral}),
using the spectral method with the boundary condition 
$\bar{\bar{\mathcal{F}}} = 0$ at $\psi=0$ and $\psi = L_{\psi} =1$. Finally 
Eq.~(\ref{QMatrix}) is solved iteratively, by calculating $\mathcal{N}_{n}$ 
for $\phi_{n}$ solution from the previous iteration. For the 1st iteration
with mode number $n$, the potential solution $\phi_{(n-\Delta n)}(\psi)$ 
of the mode $n-\Delta n$ is taken into account to calculate $\mathcal{N}_{n}$. 
But for the calculation with mode
$n=1$, at the beginning (1st iteration) the potential solution is not 
available, therefore at first, Eq.~(\ref{QMatrix}) is solved using the spectral
method for the case with $J_{0,s} =1$ and construct the potential using 
Eq.~(\ref{fmConstract}). Then the gyro-bounce average operator 
Eq.~(\ref{gyro_bounce}) is applied on that potential solution for calculating 
$\mathcal{N}_{n}$ at the $1^{st}$ iteration for $n=1$, and construct the solution 
$\phi_{n}$. In the next iteration this $\phi_{n}$ is used to calculate $\mathcal{N}_{n}$
and update the solution $\phi_n$.
After every iteration, $\Delta \phi_n$, the difference in solution $\phi_n$ from its 
value at previous iteration, is calculated and this iteration method continuous until 
achieving the accuracy limit $|\Delta\phi_n| \sim 10^{-3}$. 

Though this solver based on spectral method is quite robust for studying the 
linear stability of TIM for different dependent parameters 
(temperature, density and precession frequency) profiles, it has some 
limitations. We have observed this spectral method can calculate the 
($\gamma$, $\omega_r$) precisely for temperature profiles having 
$\kappa_T < 10 \kappa_{T,th}$, and beyond that the method becomes unstable, and
even a very small change in $\Delta n < 0.001$ departs the solution from the
expected branch with highest growth-rate.
$\kappa_{T,th}$ is the threshold value of instability which is given by 
Eq.~(\ref{KT_thr_LambdaD} and \ref{KT_thrHE}). For $H_{eq}\approx E$ and $\kappa_n = 0$ 
the threshold value is
$\kappa_{T,th} \sim 0.12$ for the mode number $n=1$ \cite{drouot14:t}. 
Moreover, this method can be applied 
only for the potential solutions having zero values at the boundary 
($\phi = 0$ at $\psi=0$ and $\psi=1$). 
In the next two sections
Sec.~\ref{compare}-\ref{LambdaD}, this spectral technique is used for studying 
the effect of temperature and precession frequency profiles 
(having $\kappa_{T,max} < 10 \kappa_{T,th}$) on TIM instability for 
the case in the limit $H_{eq}\approx E$ and for the exact equilibrium Hamiltonian 
$H_{eq}(\psi,\kappa,E)$. $\kappa_{T,max}$ is the maximum value of a $\kappa_{T}(\psi)$
profile.
%
\section{Effect of temperature and precession frequency on TIM instability in the limit
$H_{eq}\approx E$
\label{compare}}
%
In this section we consider the case with the limit $H_{eq}\approx E$ and present 
the effect of different temperature 
profiles, and different precessional frequency profiles $\Omega_D(\psi,\kappa)$, 
corresponding to different safety factor profiles, on the trapped ion mode instability.
We can substitute $\Lambda_D=1$ and 
$\kappa_{\Lambda}=0$ in the Eq.~(\ref{Ns_res_LambdaD}) and get the required expression
for $\mathcal{N}_n$ Eq.~(\ref{Ns_res_HE}), in the limit $H_{eq}\approx E$.
In this limit the growth-rate mostly depends on the terms $\kappa_T$ and $\kappa_n$ in 
Eq.~(\ref{Ns_res_LambdaD}). Moreover, for all the cases the results  
from the global linear analysis is compared with the linear TERESA simulations. 
Throughout this paper we have considered a flat density profile 
$\kappa_n = 0$, which prevents the generation of electron roots of trapped 
ion mode (TIM) ($\omega_r < 0$) with propagation along electron diamagnetic 
direction (appendix of \cite{lesur:m17}). 
Here all the modes are ion roots of TIM which propagate along the ion 
diamagnetic direction. In this sense the modes are ``{\it pure}-TIM''.
\subsection{Different temperature profiles
\label{Temp-Variation}}
\begin{figure}
\includegraphics[width=\linewidth]{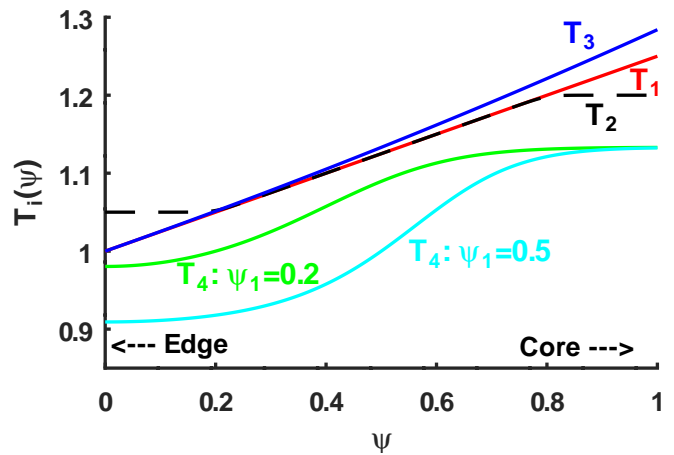}
\caption{Five different types of temperature profiles, $T_1(\psi)$ (Red solid line), 
$T_{2}(\psi)$ (black dashed line), $T_3(\psi)$ (blue solid line), $T_4(\psi)$ with 
$\psi_1 = 0.2$ (green solid line) and  $T_4(\psi)$ with $\psi_1 = 0.5$ (cyan solid line).
The radial interval in between $\psi=0$ and $\psi=1$ covers a certain radial 
domain in between the last closed fluxed surface (LCFS) and the very core 
region of a tokamak, respectively.
$\dashrightarrow$ near $\psi=1$: indicates direction towards core region of tokamak. 
$\dashleftarrow$ near $\psi = 0$: indicates direction towards edge region 
(last closed flux surface) of a tokamak.
\label{fig_Tprof}}
\end{figure}

In this sec.~(\ref{Temp-Variation}) we use four 
different types of temperature profiles and a flat density profile $\kappa_n=0$, and
investigate their effect on the trapped ion mode (TIM) instability.
These temperature profiles are presented as,
\begin{eqnarray}
\begin{aligned}
	T_1(\psi) &= T_0\left(1+ G_{T} \psi\right) \\
	T_2(\psi) &= T_0 + \frac{G_{T}}{2}\Bigg(\Bigg.1+L_{1}\Bigg[\Bigg.
		\log\left(\cosh\frac{\psi -\psi_{1}}{L_{1}}\right) \\
	&- \log\left(\cosh\frac{\psi -(1-\psi_{1})}{L_{1}}\right)
	\Bigg.\Bigg]\Bigg.\Bigg)\\
	T_3(\psi)  &= T_0 \exp\left(G_{T} \psi\right) \\
	T_{4}(\psi) &= T_0 \exp\left[ G_{T} R_{T} \tanh\frac{\psi^2-\psi_1^2}
	{R_T^2}\right]
\end{aligned}
\label{T0_Profie}
\end{eqnarray}
Here we use the constant parameters $T_0 = 1$, $G_T =0.25$, 
$L_{1} = 0.025$, 
$\psi_1 = 0.2$, $R_T = 0.5$ and for the temperature profile $T_4$ we have 
chosen two different values of $\psi_1$, ($0.2$ and $0.5$). 
Fig.~\ref{fig_Tprof} presents all these five different temperature profiles.
$T_1(\psi)$ is the temperature profile 
with constant temperature gradient. $T_2(\psi)$ has constant gradient inside
the simulation domain from $0.2\le\psi\le0.8$. The width of this region is 
controlled by the parameter $\psi_1$ in $T_2$. Near the boundary, the gradient is 
zero along the $\psi$ direction. Therefore there is a sudden change in gradient 
at $\psi=0.2$ and $0.8$. This type of profile with zero gradient at boundary
is used in TERESA 
simulation to prevent numerical instabilities arising from the boundaries. $T_3(\psi)$
is an exponentially increasing temperature profile. $T_4(\psi)$ is a special
type of temperature profile which has zero gradient near boundary, and inside 
the box it has finite gradient which can be controlled by the parameter $\psi_1$ 
keeping $G_T$ constant. Here two different values of $\psi_1$ are used $\psi_1 = 0.2$ and
$\psi_1 = 0.5$.
For both values of $\psi_1$ the gradient changes smoothly from zero to 
finite value as $\psi$ goes from boundary towards center region. Therefore this profile 
has less-chance to generate numerical instabilities. 
Temperature $T_i$ is normalized to $T_0$, and the poloidal magnetic flux is 
normalized such that $\psi = 0$ is close to last-closed flux surface and
$\psi = 1$ is in the core region of a tokamak.
Therefore $T_i = 0.9$ is associated with the ion temperature  $0.9 T_0$. Moreover, 
for any particular temperature profile, as an example $T_1$, the core-most 
temperature ($\psi = 1$) is $1.25 T_0$ and the temperature
at edge ($\psi = 0$) is $T_0$, so that $\psi\in [0, 1]$ corresponds to a limited radial 
extent of a tokamak where the temperature changes by only $25\%$.
In tokamak experiments the ion 
temperature profile depends on operation-mode, during L-mode operation it looks like 
a bell shaped profile with a flat top near the core and decreases towards the 
Scrape-Off Layer
region, and during H-mode operation due to generation of an edge transport barrier the 
plasma temperature profile changes significantly. It is difficult to present 
the entire temperature profile using a single function, but the profiles $T_1$, $T_2$, 
$T_3$ and $T_4$ are somewhat representative of radial sections as we have checked from
different tokamak experiments e.g. data from JET \cite{ottaviani97:m, jacquinto:j},
COMPASS \cite{panek:r}.

\begin{figure}
\includegraphics[width=\linewidth]{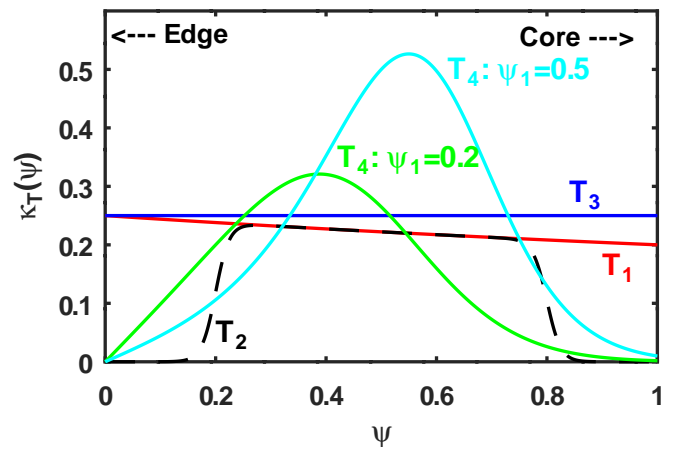}
\caption{$\kappa_T = \frac{d \log T(\psi)}{d\psi}$ profiles for the five 
different temperature profiles. 
\label{fig_KTprof}}
\end{figure}

Fig.~\ref{fig_KTprof} presents the 
$ \kappa_T(\psi) = \frac{1}{T(\psi)} \frac{dT}{d\psi}$ profiles for all these five 
temperature profiles. The $\kappa_T$ profiles change significantly for all these 
five temperature profiles. 
The temperature profile $T_4$ with $\psi_1=0.5$ and $0.2$ have
$\kappa_T$ profiles with peak values $ 0.53$ at $\psi\approx 0.6$
and $0.31$ at $\psi\approx 0.4$, respectively and then gradually 
decreases to zero towards the boundaries.
The exponential temperature profile $T_3$ has constant $\kappa_T$. 
Profiles $T_1$ and $T_2$ have exactly 
similar $\kappa_T$ profiles, $\kappa_T \propto \psi^{-1}$ within the region 
$0.2\le\psi\le 0.8$, 
and near the boundaries $\psi < 0.2$ and $\psi > 0.8$, $\kappa_T$ suddenly jumps to zero 
for $T_2$. Therefore in $T_2$ the TIM instability arises
due to the temperature profile within $0.2\le\psi\le 0.8$. 
%
\begin{table}[h!]
\centering
\begin{tabular}{ c c c c c c c c c}
\hline
~$\delta_{b,i}$~ &~ $\rho_{c,i}$~ &~ $C_{ad}$~ &~ $C_{pol}$~ &~ $\Omega_{D}$~ 
&~$T_{eq}$~ &~ $n_{eq}$~ &~ $M$~ & ~$G_{T}$~\\
\hline
~0.1~ &~ 0.03~ &~ 0.1~ &~ 0.1~ &~ 1~ &~ 1~ &~ 1~ &~ 50~&~0.25~\\
\hline
\end{tabular}
\caption{Main input parameters used for studies in sec.~\ref{Temp-Variation}.}
\label{table:1}
\end{table}
\subsubsection{$J_{0} = 1$ Case
\label{J0_1}}
Since the case with gyro-bounce
average operator $J_0 =1$ is numerically the simplest case, we first validate our
global linear analysis for the case with $J_0 =1$ and with five different 
temperature profiles. 
Moreover for mode number $n=1$, the potential solution 
from the $J_0=1$ case will be used to calculate $\mathcal{N}_n$.
The other main parameters which are used in this study are presented in 
the table~\ref{table:1}. 
\begin{figure}
\includegraphics[width=\linewidth]{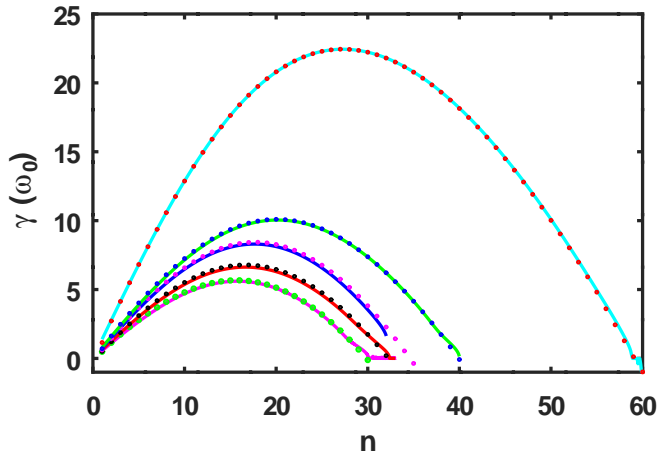}
\caption{Growth rate $\gamma$ Vs. mode number $n$ profiles
with $J_0 =1$, for the five different temperature profiles, $T_4(\psi)$ with 
$\psi_1 =0.5$ (cyan line), $T_4(\psi)$ with $\psi_1=0.2$ (green line), $T_3(\psi)$ 
(blue line), $T_1(\psi)$ (red line) and $T_2(\psi)$ (magenta line). The dots
denotes the results from the linear TERESA simulation with $J_0=1$. Each set of
dots marked by same color is from a single simulation. 
\label{Growth_Tpr_J01}}
\end{figure}
%
\begin{figure}
\includegraphics[width=\linewidth]{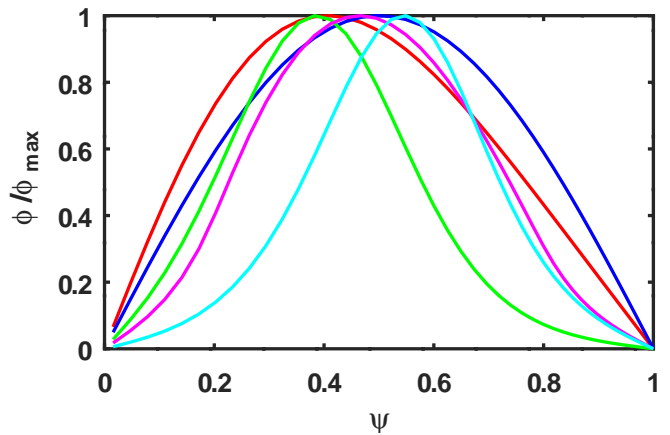}
\caption{Potential solutions $|\phi_n(\psi)|/\phi_{max}$ for mode $n=1$ with
        $J_0=1$ for the temperature profiles
        $T_4(\psi)$ with $\psi_1 =0.5$ (cyan line), $T_4(\psi)$ with $\psi_1 =0.2$
        (green line), $T_1(\psi)$ (red line), $T_2(\psi)$ (magenta line), and
        $T_3(\psi)$ (blue line).
\label{phi_J01}}
\end{figure}
%
The growth rate $\gamma$ of TIM instability for different mode 
numbers $n$ is presented in Fig.~\ref{Growth_Tpr_J01} for all the five 
temperature profiles. This spectral method is also able to find out the 
negative growth-rates for the higher mode numbers. 
Since we are interested only in the unstable TIM modes
with positive growth-rates ($\gamma>0$), in all the figures of $\gamma-n$ 
profiles, only the modes with $\gamma>0$ are presented.
For the temperature profile $T_4$ with $\psi_1=0.5$, the effective 
temperature gradient and hence the parameter $\kappa_T$ has the highest value 
compared to 
other temperature profiles. Therefore as expected the growth-rate of the TIM instability 
is higher compared to the other cases. For the other temperature profiles the 
growth-rate decreases as the effective value of $\kappa_T$ decreases. 
The growth-rate of the highest growing mode for the temperature profiles $T_4$ with 
$\psi_1 = 0.5$, $T_4$ with $\psi_1 = 0.2$, $T_3$, $T_2$ and $T_1$ are $22.5$, $10$, 
$8.3$, $5.6$ and $6.6 \omega_{0}$, respectively, and the location of most unstable modes 
are $n\approx 27$, $20$, $17$, $15$ and $16$, respectively.
Therefore even for the gyro average operator $J_0 = 1$ case the 
growth rate of the TIM instability strongly depends on the temperature profile,
and the mode with highest growth rate shifts towards zero with decreasing
value of $\kappa_T$. For low mode numbers the growth-rate increases 
with increasing $n$. At very high $n$ (after the mode with highest growth 
rate) the growth-rate decreases and goes down to zero due to the presence of 
polarization drift term within the model, and 
finally generates a bell shaped growth-rate profile. Moreover, the presence of 
gyro-average operator $J_0$ according to Eq.~(\ref{gyro_bounce}), will enhance this 
effect that decreases the growth-rate further for higher mode numbers, which is 
presented in the next sec.~\ref{J0_Pade}.  
All the growth-rate profiles at large values of mode-number $n$, where 
$\gamma \rightarrow 0$ and $\omega_r$ has very large values ($\omega_r\gg  \gamma$),
slightly depart from expected trend. The reason is that,
the integration of Eq.~(\ref{Ns_res_LambdaD}) for calculating 
$\mathcal{N}_n$ is a contour integration, and in the complex plane the contour/path 
of the integration depends on the sign of imaginary part of $\chi$ ie., $\gamma$. 
Since we are interested in the modes with $\gamma > 0$, we have
considered the integration path for the pole at $E = C(\omega_r + i\gamma)$, where
$C = (n\Omega_D)^{-1}$. For negative values of $\gamma$ and $\gamma = 0$ 
the contours of the integration will be different, because 
the poles are situated at $E = C(\omega_r - i\gamma)$ and $E = C\omega_r$, respectively. 
One has to 
take into account these new contours for getting the exact values of growthrate 
when $\gamma \rightarrow 0$ and $\gamma \le 0$. However for higher $\omega_r$ the 
phase velocity of the wave is very high, and the kinetic effect of trapped ions to 
TIM modes are negligible. Therefore the growthrate of those modes is left out of 
our present study. 
The dots on the growth-rate curves present the measured 
growth-rates of the different modes $n$ for different temperature profiles 
from the linear TERESA simulation with $J_0 =1$. 
These good agreements of the
growth-rate profiles validate the accuracy of our spectral 
method to the semi-Lagrangian based Vlasov simulation results.  
For generating an entire ($\gamma$ Vs. $n$) curve, our spectral method based solver 
takes only $10-15$ minutes for $J_0=1$ and $20-30$ minutes for $J_0$ according to 
Pad\'e expression, whereas a serial version of TERESA simulation takes around
$16$ hours. Fig.~\ref{phi_J01} presents the potential solution $\phi_n(\psi)$ profiles
for $n=1$ for the different temperature profiles. Depending on the $\kappa_T$ profiles
different $\phi_n(\psi)$ profiles are generated. 
Since the potential solution has both real and imaginary parts, for all the potential 
profiles within this paper, we 
take its absolute values and then normalize by its maximum value.   
All the results that are presented hereafter use the gyro-bounce average operator 
according to the Pad\'e expression Eq.~(\ref{gyro_bounce})
\subsubsection{$J_{0}$ according to Pad\'e expression 
\label{J0_Pade}}
\begin{figure}
\includegraphics[width=\linewidth]{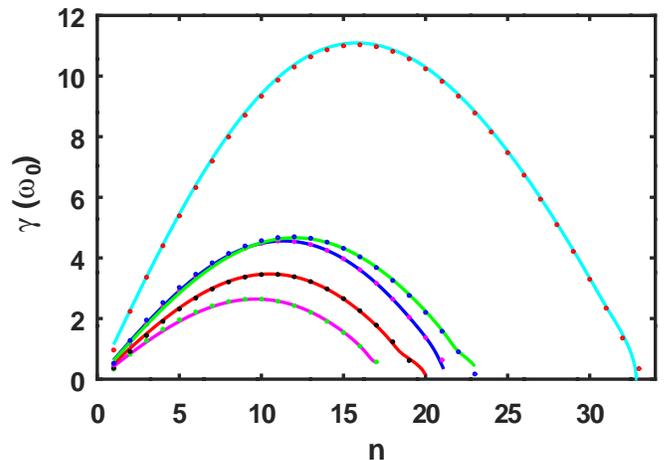}
\caption{ $\gamma$ Vs. $n$ profiles with 
$J_0$ Eq.~(\ref{gyro_bounce}) for the five different temperature profiles, 
$T_4(\psi)$ with $\psi_1 =0.5$ (cyan line), and $T_4(\psi)$ with $\psi_1=0.2$ 
(green line), $T_3(\psi)$ (blue line), $T_1(\psi)$ (red line) and $T_2(\psi)$ 
(magenta line). The dots denote the results from the linear TERESA simulation. 
\label{Growth_Tpr_J0Pade}}
\end{figure}

After validating the $J_0 = 1$ case, we intend to solve the differential 
equation Eq.~(\ref{QMatrix}) by
calculating $\mathcal{N}_n$ from Eq.~(\ref{Ns_res_HE}) with the Pad\'e 
expression for the gyro-bounce average operator $J_0$ Eq.~(\ref{gyro_bounce}).
The gyro-bounce averaged quantity ($J_0 f_{n}$ and $J_0 \phi_n$) is calculated by 
solving the differential 
Eq.~(\ref{Gyro_averageSpectral}) using the spectral method. First 
$J_{0,n} \phi_{n,\omega}$ is calculated using the solution $\phi_n$, calculated
from Eq.~(\ref{QMatrix}) by implementing $J_0 = 1$. Then we use this 
gyro-bounce averaged potential $\bar{\phi} = J_{0,n} \phi_{n,\omega}$ 
in Eq.~(\ref{linearfSol_HE}) to construct the particle distribution  $f_n$, and 
finally the gyro-bounce averaged distribution function $\bar{f_n}$ is obtained 
similarly 
by solving the differential Eq.~(\ref{Gyro_averageSpectral}) which is used to 
calculate $\mathcal{N}_n$. Fig.~\ref{Growth_Tpr_J0Pade} presents
the growth-rate of different modes $n$ of TIM instability, for the five different 
temperatures profiles Eq.~(\ref{T0_Profie}). 
In comparison with the previous $J_0=1$ cases, 
the gyro-bounce averaged operator decreases the growth rate $\gamma$ of the TIM instability 
by a significant amount for all the temperature profiles, roughly by a factor 
two in terms of highest growth rate. Also the modes with the 
highest growth rate for each temperature profile are different than the previous 
$J_0=1$ case and they are shifted towards lower mode numbers.
%
Therefore the gyro-average operator reduces the instability of the TIM modes
compared to $J_0 = 1$  case. But the effect of the variation in temperature 
profiles on their growth rate remains unchanged. Similar to previous $J_0=1$ 
case, the growth rate of TIM modes depends on $\kappa_{T}$ value, and the 
profile $T_4$ with $\psi_1 = 0.5$ generates the highest growth-rate, which decreases
as $\kappa_{T}$ value decreases for the other temperature profiles.
Each dots with different colors on the
solid lines present the growth-rate of different modes $n$ for different
temperature profiles from the linear TERESA simulation using Pad\'e expression for 
gyro-bounce average operator. 
\begin{figure}
\includegraphics[width=\linewidth]{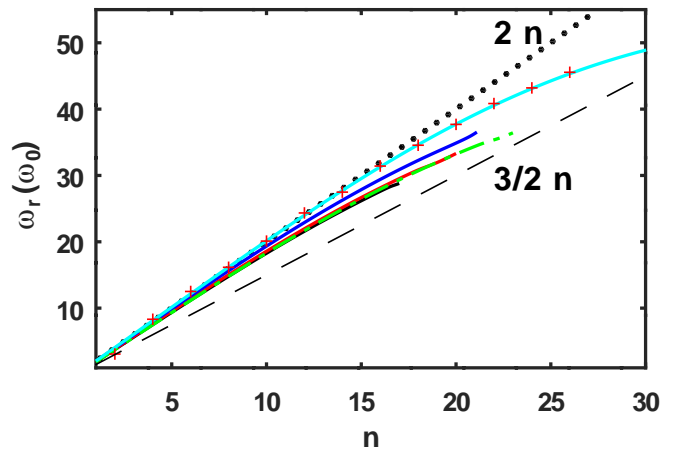}
\caption{Real part of TIM frequency $\omega_r$ Vs. mode-number $n$ with 
        $J_0$ Eq.~(\ref{gyro_bounce}), for five different temperature profiles,
	$T_4(\psi)$ with $\psi_1 =0.5$ (cyan line), $T_4(\psi)$ with 
	$\psi_1=0.2$ (dashed green line), $T_2(\psi)$ (black line), 
	$T_1(\psi)$ (red line) and 
	$T_3(\psi)$ (blue line). The red$+$ marker presents the linear TERESA 
	simulation results for $T_4$ with $\psi_1= 0.5$. 
	The dashed black line presents $\omega_r = 3/2 n$, and the black doted 
	line presents $\omega_r = 2 n$
\label{Omegar_Tpr_J0Pade}}
\end{figure}
%
Fig.~\ref{Omegar_Tpr_J0Pade} presents the frequency $\omega_r$ (real part) of 
different trapped ion modes (TIMs) $n$ for five different temperature profiles.
For the lower mode numbers $n \le 10$ frequencies
$\omega_r$ are almost similar for all the temperature profiles.
The black dashed line presents $\omega_r = \frac{3}{2} n$ which is the 
solution of the local linear analysis Eq.~(\ref{Loc_linHE}) in the limit of 
$\psi \rightarrow 0$ \cite{drouot14:t}.
Therefore the frequencies $\omega_r$ of TIMs are higher compared to the local
linear case. Indeed we found for these cases they follow 
$\omega_r \sim 2 n$ relation for the global linear analysis. The red
marker `$+$' presents the results from the linear TERESA simulation for the
temperature profile $T_4$ with $\psi_1=0.5$.
\begin{figure}
\includegraphics[width=\linewidth]{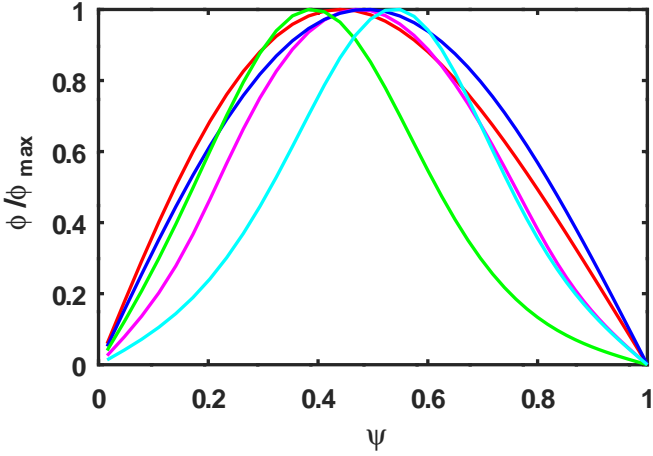}
	\caption{Potential solutions $|\phi_n(\psi)|/\phi_{max}$ for mode $n=1$ with 
        $J_0$ Eq.~(\ref{gyro_bounce}) for the five different temperature profiles
	$T_4(\psi)$ with $\psi_1 =0.5$ (cyan line), $T_4(\psi)$ with $\psi_1 =0.2$
	(green line), $T_1(\psi)$ (red line), $T_2(\psi)$ (magenta line), and 
	$T_3(\psi)$ (blue line).
\label{Phi_J0Pade}}
\end{figure}
%
Fig.~\ref{Phi_J0Pade} presents the potential solutions $\phi_n(\psi)$ of the 
Eq.~(\ref{QMatrix}) for $n=1$, for all the five different temperature profiles.
The potential profiles are slightly different from the 
previous case with $J_0=1$ (Fig.~\ref{phi_J01}), due to the gyro-bounce average operator 
$J_0$ Eq.~(\ref{gyro_bounce}). 
\begin{figure}
\includegraphics[width=\linewidth]{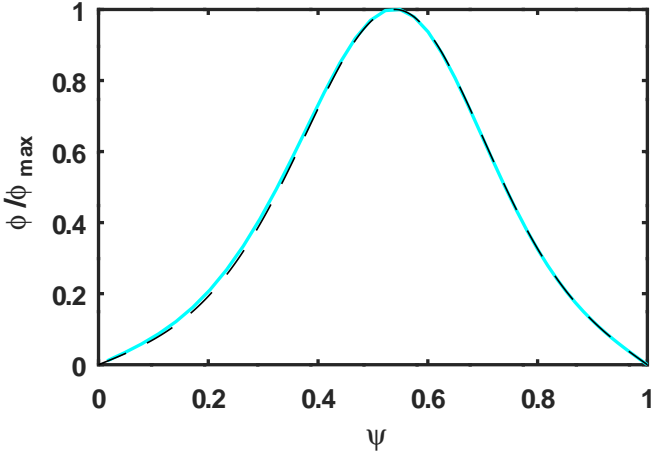}
\caption{Potential solution for the temperature profile $T_4$ with 
$\psi_{1}= 0.5$ for the highest growing mode $n = 15$. Black dashed line 
presents the potential solution from the TERESA linear simulation.
\label{Phi_comp_J0PadeT4}}
\end{figure}
%
\begin{figure}
\includegraphics[width=\linewidth]{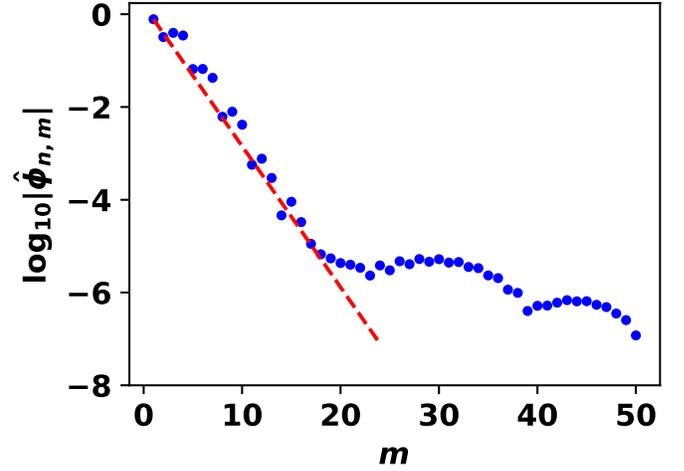}
	\caption{Coefficients $\log{(|\hat{\phi}_{n,m}|)}$ of the function 
	$\mathcal{C}_m$ vs. $m$ for $M=50$ points along $\psi$ in a computation for the mode $n=15$ 
in case of temperature profile $T_4$ with $\psi_{1}= 0.5$. 
A function $\log[A_{0} \exp(-\beta m)]$ is plotted for $\beta=0.7$ 
(red dashed line).
$A_0$ is a constant.
\label{am_coeff}}
\end{figure}
%
Fig.~\ref{Phi_comp_J0PadeT4} shows the $\phi_n(\psi)$ 
solution profile Eq.~(\ref{fmConstract}) of mode number $n = 15$ for the 
temperature $T_4$ with 
$\psi_1=0.5$, which is almost similar to the mode number $n=1$ (cyan line in 
fig.~\ref{Phi_J0Pade}). Dashed black line presents the potential profile from 
the linear TERESA simulation for the temperature profile $T_4$ with $\psi_1=0.5$ 
and mode number
$n = 15$. The potential solution is also in good agreement with the 
TERESA simulation. Fig.~\ref{am_coeff} presents the decimal logarithms of the 
coefficients $\hat{\phi}_{n,m}$ of the function $\mathcal{C}_m$ in 
Eq.~(\ref{fmConstract}) for 
the mode number $n=15$ for the case with temperature profile $T_4$($\psi_1=0.5$). 
The coefficients $\hat{\phi}_{n,m}$ decrease as 
$m$ increases. This confirms the spectral convergence of $\hat{\phi}_{n,m}$, 
which follows
$|\hat{\phi}_{n,m}| \sim A \exp(-\beta m)$, with $\beta = 0.7$ for $1\le m\le 18$ 
(red dashed line in fig.~\ref{am_coeff}). 
Note that $|\hat{\phi}_{n,m}/\hat{\phi}_{n,1}|< 10^{-5}$ for $m>18$ which suggests that
$M=18$ is sufficient for generating the results with good accuracy in this case. 
However in the next section the profile for $\Omega_D$ is also included, which 
changes the coefficients $\hat{\phi}_{n,m}$ values. Therefore to confirm the good accuracy 
for all other cases we consider $M=50$ throughout this manuscript.
Hereafter we will consider only the linear temperature 
profile $T_1$ and vary other important parameters for TIM instability.
\subsection{Variation in precession frequency 
\label{Preces-Variation}}
According to Eqs.~(\ref{omegaD},\ref{omegabar}), the precession frequency
$\Omega_D$ depends on both $\psi$ and $\kappa$. In the previous section we 
consider a constant value of $\Omega_D = 1$. In this section we will first 
consider the effect of trapping parameter $\kappa$ on the precession 
frequency, and second, consider the $\psi$ dependency of $\Omega_D$ and 
investigate their effects on the TIM instability in the limit of $H_{eq}\approx E$. 
\begin{table}[h!]
\centering
\begin{tabular}{ c c c c c c c c c}
\hline
	~$\delta_{b,i}$~ &~ $\rho_{c,i}$~ &~ $C_{ad}$~ &~ $C_{pol}$~ &~ $T(\psi)$~ 
&~$T_{eq}$~ &~ $n_{eq}$~ &~ $M$~ & ~$G_{T}$~\\
\hline
	~0.1~ &~ 0.03~ &~ 0.1~ &~ 0.1~ &~ $T_{1}$~ &~ 1~ &~ 1~ &~ 50~&~0.25~\\
\hline
\end{tabular}
\caption{Main input parameters used for studies in sec.~\ref{Preces-Variation}}
\label{table:2}
\end{table}
\subsubsection{$\kappa$ dependency of $\Omega_D$ 
\label{Kappa-ver-OmegaD}}
%
In a banana orbit the critical poloidal angle $\theta_{crit}$, where 
$v_{\parallel}=0$, is linked with the trapping parameter $\kappa$ as 
$\kappa^2 = \sin^2\left(\frac{\theta_{crit}}{2}\right)$. Therefore the 
acceptable values of $\kappa$ is $0\le\kappa\le 1$. $\kappa= 0$ is associated 
with the particles having almost zero parallel velocity and therefore their 
motions are restricted close to the center of the banana and are called deeply 
trapped particles. $\kappa \sim 1$ is associated with the particles having 
maximum parallel velocity for the trapped particles, 
$\frac{|v_{\parallel}|}{|v_{\perp}|} \sim \sqrt{\frac{B_{max}}{B_{min}}-1}$,
whose reflecting point is situated near the position of maximum magnetic field 
$B_{max}$ at innerside of tokamak, and are called barely trapped particles. 
Eq.~(\ref{omegabar}) presents the theoretical $\kappa$ dependency of the 
precession frequency $\bar{\omega}_d(\kappa)$. 
\begin{figure}
\includegraphics[width=\linewidth]{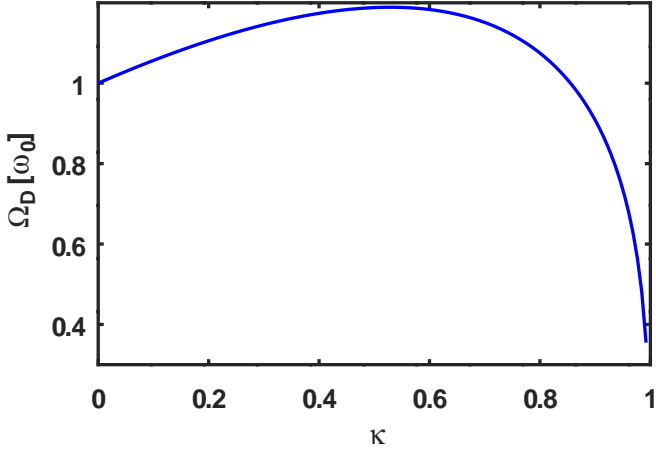}
	\caption{Precession frequency $\bar{\omega}_d(\kappa)$ Eq.~(\ref{omegabar}) for 
magnetic-shear $s_0 = 0.8$. 
\label{OmegsD_Kappa}}
\end{figure}
%
Fig.~\ref{OmegsD_Kappa} presents this 
variation of $\bar{\omega}_d$ with $\kappa$ for a constant 
value of magnetic-shear $s_0=0.8$. In absence of $\psi$ dependency of 
precession frequency, we can write $\Omega_D(\kappa) =\bar{\omega}_d$ from
Eq.~(\ref{omegaD}). For a realistic safety factor profile \cite{gobbin:m},
the magnetic shear $s_0(\psi)$ decreases almost
linearly in $\psi$, and at $\psi=0.5$ it has value $s_0\sim 0.8$. In previous 
section-(\ref{Temp-Variation}), we considered deeply trapped particles with 
$\kappa=0$. Here we consider two different cases with same temperature 
profile $T_1$. In one case we consider constant value of precession frequency
$\Omega_D = 0.6$ associated with barely trapped particles ($\kappa \sim 1$), 
and to compare the results we reconsider the previous case with 
$\Omega_D = 1$ ($\kappa=0$)
and temperature profile $T_1$. In second case to incorporate the effect of 
{\it pitch-angle} dependency in the TIM-instability we consider the 
entire $\Omega_D(\kappa)$ profile for the trapped particles. In this 2nd 
case the dispersion relation is calculated by doing the integration along
 $\kappa$ systemically according to Eq.~(\ref{quasi-Nutr2}), which was 
neglected for constant $\Omega_D$ by replacing 
$\int_0^1 \kappa \mathcal{K}(\kappa^2) d\kappa = 1$.
One important thing is that, for both cases we assume that the fraction of 
trapped particle $f_t \sim \sqrt{2\varepsilon}$ remains 
constant, which is controlled by the parameter 
$C_{ad}$ in Eq.~(\ref{quasi-Nutr2}). If we consider radial dependency of inverse
of aspect ratio as $\varepsilon = \frac{r}{R_0}$, the fraction of trapped 
particle can be written as $f_t \sim (1-\psi)^{1/4}$. This justifies
our approximation $f_t\sim const.$ throughout the region $\psi\in [0,1]$.
Therefore in 1st case with 
constant $\Omega_D=0.6$, all trapped-particles are trapped near the separatrix 
(barely trapped), and for $\Omega_D=1$, equal amount of particles are trapped
near the center of banana, whereas in the 2nd case with entire 
$\Omega_D(\kappa)$ profile, equal amount of trapped particles are distributed 
over the entire trapped domain.
The other essential parameters are taken from Tab.~\ref{table:2}.
\begin{figure}
\includegraphics[width=\linewidth]{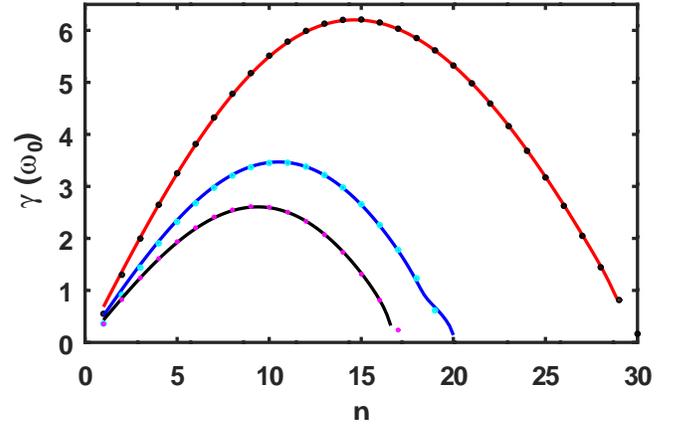}
\caption{Growth rate $\gamma$ Vs. mode-number $n$ with $J_0$ from
Eq.~(\ref{gyro_bounce}) and temperature $T_1(\psi)$, for constant values of 
precession frequency $\Omega_D = 1$ (solid blue line) and $0.6$ 
(solid red line). The black solid line is for entire profile of 
$\Omega_D(\kappa)$ Eq.~(\ref{omegabar}). Dots are the results from the linear 
TERESA simulations.
\label{Growth_OmegaD_kapa}}
\end{figure}
%

Fig.~\ref{Growth_OmegaD_kapa} presents the growth-rates $\gamma$ for 
different values of mode numbers $n$ of TIM-instability for these cases.
Since the TIM instability occurs due to the resonance of precession frequency 
$\Omega_D$ of particles with the wave frequency $\omega$, 
for low precession frequency $\Omega_D = 0.6$ 
the wave with smaller phase velocity (smaller $\omega$) resonates. 
In the equilibrium distribution $F_{eq}(E)$ Eq.~(\ref{Fequilibrium_HE}), there 
are large number of particles near lower velocity 
($E \equiv \frac{1}{2} m v_{G\parallel}^2 + \mu B_{G}$) compared to higher
velocity, as a consequence the charge separation due to $\nabla B$ drift in
presence of temperature gradient would be higher near the lower velocity 
compared to higher velocity (higher $\Omega_D$), and
generates stronger electric field which helps to enhance the density 
perturbation of the wave. 
Therefore $\Omega_D = 0.6$ has higher growth-rate compared to the case with 
higher precession frequency $\Omega_D = 1$. 
Since the real part of the TIM frequency $\omega_r \propto \Omega_D$ 
Eq.~(\ref{Loc_linHE}), for lower $\Omega_D = 0.6$ value $\omega_r$ is 
significantly smaller compared to $\Omega_D =1$ case.
If we consider the entire profile of $\Omega_{D}(\kappa)$ for the constant 
magnetic-shear $s_0=0.8$ (Fig.~\ref{OmegsD_Kappa}), up to $\kappa \le 0.85$, 
the precession frequency 
$\Omega_D \ge 1$, and within $0.85\le \kappa \le 1$ it has smaller value
$\Omega_D < 1$. Since the equilibrium distribution $F_{eq}$ Eq.~(\ref{Fequilibrium_HE}) 
is independent of $\kappa$, there are equal number of particles at all $\kappa$
values. Therefore in this case, most of the particles (almost $80\%$) have precession 
frequency $\Omega_D > 1$, only $20\%$ particles have precession 
frequency $\Omega_D < 1$. As a result the contribution from the $\Omega_D > 1$
dominates, therefore the waves with higher phase velocity (higher $\omega$) 
resonates with the particles, where less number of particles are available due to the
Maxwellian particle distribution in $v_{\parallel}$. This decreases the growth-rate 
$\gamma$ compared to the $\Omega_D = 1$ case. 
The entire $\kappa$ dependent $\Omega_{D}(\kappa)$
profile decreases the growth-rate by $30 - 50 \%$ compared to constant 
$\Omega_D$ value at a constant $\kappa$ location.
Dots are the growth-rates
for different mode numbers $n$ from linear TERESA simulation.
\subsubsection{$\psi$ dependency of $\Omega_D$ 
\label{Psi-ver-OmegaD}}
The theoretical dependency of precession frequency on radius $r$ is given by
Eqs.~(\ref{omegaD}-\ref{omegabar}) where both the safety factor $q(r)$, and
magnetic shear $s_0(r) = \frac{r}{q(r)}\frac{dq}{dr}$ depends on $r$. For a 
particular $q(r)$ profile $q(r) = 1.1 +2 r^2$ profile \cite{gobbin:m}, 
the magnetic flux-function is calculated from the integration 
$\psi(r)\sim -B_{min}\int_{r_{in}}^{a} \frac{r}{q(r)}dr$, 
and normalized according to Tab.~\ref{table:normaliz}, where 
$L_{\psi} = |\psi(r_{in})-\psi(a)|$ is the length of the simulation box in $\psi$ unit.
Finally $r$ depends on $\psi$ as, 
\begin{eqnarray}
	r = \sqrt{\frac{q(a)}{2}\left(\frac{q(r_{in})}{q(a)}\right)^{\psi} 
	- \frac{q(0)}{2}},
\label{Psi_r}
\end{eqnarray}
where $q(a)$, $q(0)$ and $q(r_{in})$ are the value of safety factor at $r=a$, 
$r=0$ and $r = r_{in}$, where $a$ is the minor radius of tokamak and $r_{in}$ 
is the lower limit of $r$ integration in $\psi(r)$ expression, which helps 
to remove the singular nature of $\omega_d$ in Eq.~(\ref{omegaD}) at 
$r\rightarrow 0$.
\begin{figure}
\includegraphics[width=\linewidth]{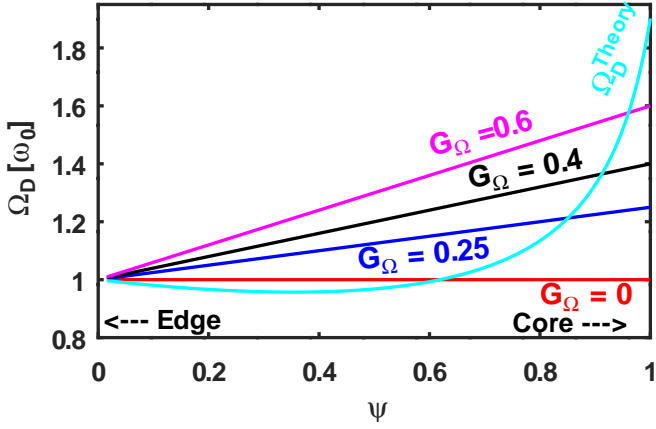}
\caption{Precession frequency profiles
$\Omega_D(\psi) = \Omega_{D0} + G_{\Omega}\psi$, with $G_{\Omega} = 0$ (solid 
red line), $G_{\Omega} = 0.25$ (solid blue line), $G_{\Omega} = 0.4$ (solid 
black line) and $G_{\Omega} = 0.6$ (solid magenta line). $\Omega_{D0} = 1$ for 
all the cases. The theoretical $\Omega_D(\psi)$ profile (solid cyan line) is 
calculated from Eq.~(\ref{omegaD}) for $\kappa=0$ and $q(r) = 1.1+2r^2$.
\label{FigOmegaD_Psi}}
\end{figure}
Using this expression of $r$ Eq.~(\ref{Psi_r}) into $q(r)$ and $s_0(r)$, the 
$\psi$ dependent safety factor
$q(\psi)$ and magnetic shear $s_0(\psi)$ can be obtained, and finally 
$\Omega_D(\psi,\kappa)$ can be calculated from Eqs.~(\ref{omegaD}-\ref{omegabar}).
The theoretical $\Omega_D(\psi)$ profile for $r_{in} = 0.2$ and $\kappa=0$ is 
presented by solid 
cyan line in Fig.~\ref{FigOmegaD_Psi}, which has almost constant value 
$\Omega_D \sim 1$ within $0\le\psi\le 0.75$ and after that near the core region 
of tokamak ($\psi\rightarrow 1$), $\Omega_D$ increases abruptly with increase 
in $\psi$. For simplicity, first we consider $\Omega_D$ varies linearly with 
$\psi$ as:
\begin{eqnarray}
	\Omega_D(\psi) = \Omega_{0}+ G_{\Omega}\psi,
\label{OmegaD_Psi}
\end{eqnarray}
and study the effect of $\psi$ dependent $\Omega_D(\psi)$ on TIM instability.
Here $\Omega_{0}=1$ is the value of $\Omega_D$ at $\psi=0$, and $G_{\Omega}$ 
is the gradient in the $\Omega_D(\psi)$ profile. 
This linearly increasing $\Omega_D(\psi)$ profile for different values of
$G_{\Omega} = 0$, $0.25$, $0.4$ and $0.6$ are presented in 
Fig.~\ref{FigOmegaD_Psi}.
In this section the temperature profile $T_1(\psi)$ with $G_T = 0.25$, 
and $\Omega_D(\psi)$ profile with $G_{\Omega} = 0.25$ and $G_{\Omega} = 0$ 
Eq.~(\ref{OmegaD_Psi})
are taken into account. The other $\Omega_D(\psi)$ profiles will be considered in
the next section Sec.~\ref{LambdaD}.
The other essential parameters are taken from the Tab.~\ref{table:2}.
\begin{figure}
\includegraphics[width=\linewidth]{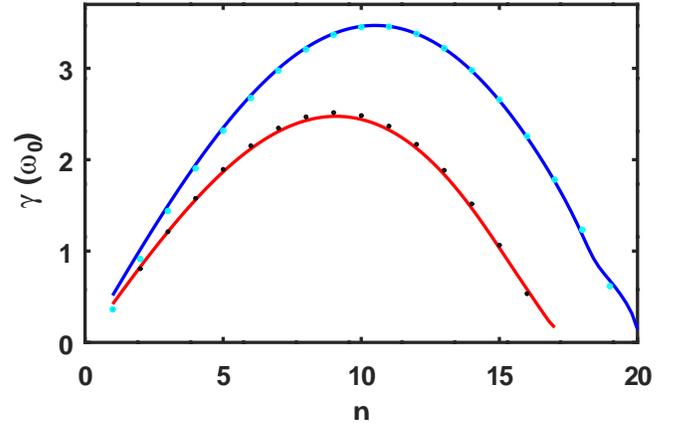}
\caption{Growth rate $\gamma$ Vs. mode-number $n$ with $J_0$ from 
Eq.~(\ref{gyro_bounce}) and temperature $T_1(\psi)$, and $\psi$ dependent 
precession frequency $\Omega_D(\psi) = \Omega_{0}+0.25\psi$ is presented in 
solid red line. Solid blue line presents the $\Omega_D =1$ case. Dots are the 
results from the linear TERESA simulations.
\label{Growth_OmegaD_Psi}}
\end{figure}
%
\begin{figure}
\includegraphics[width=\linewidth]{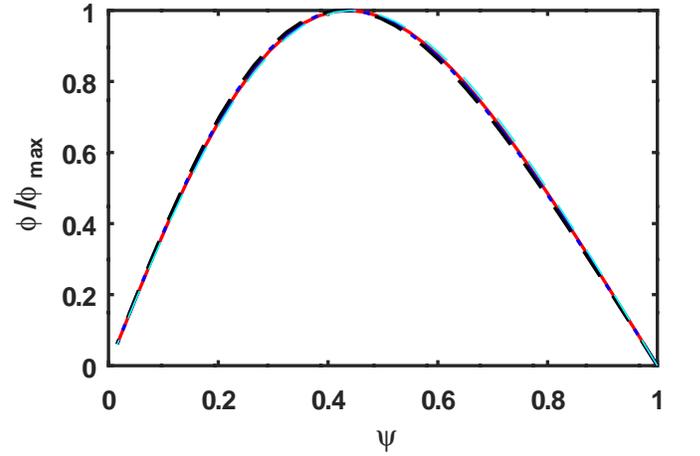}
\caption{Potential solutions $|\phi_n(\psi)|/\phi_{max}$ for mode $n=1$ with
	$J_0$ Eq.~(\ref{gyro_bounce}) for the precession frequency $\Omega_D=1$ 
	(dashed cyan), $\Omega_D=0.6$ (solid red), $\Omega_D=1+0.25\psi$ 
	(dashed black), $\Omega_D(\kappa)$ (dashed blue). 
\label{phi_OmegaD_psi_kappa}}
\end{figure}
%

Fig.~\ref{Growth_OmegaD_Psi} presents the growth-rates of
TIM modes $n$, for the $\Omega_D(\psi)$ profile 
Eq.~(\ref{OmegaD_Psi}) with $G_{\Omega} = 0.25$, and $G_{\Omega} = 0$.
As discussed in the previous sub-section (\ref{Kappa-ver-OmegaD}), 
for larger $\Omega_D$, the number of resonant particles decreases compared to 
the case with smaller $\Omega_D$ value, and as a consequence generates smaller
growth-rate of the TIM instability.
Since in the $\Omega_D$ profile with $G_{\Omega}=0.25$ 
Eq.~(\ref{OmegaD_Psi}), for all $\psi$ value 
$\Omega_D \ge 1$ and the temperature profile $T_1(\psi)$ remains unchanged,
the growth-rate $\gamma$ 
for all $n$ values is smaller compared to the case with $\Omega_D = 1$.
Here the dots present the results from the 
linear TERESA simulation.
The real part of the frequency $\omega_r$ for the TIM instability is 
proportional to $\Omega_D$ Eq.~(\ref{Loc_linHE}). Therefore in this case 
$\omega_r$ will be larger compared to the
case with $\Omega_D = 1$. However we have not presented the $\omega_r -n$ profile
within this manuscript.
Fig.~\ref{phi_OmegaD_psi_kappa} presents the potential $\phi_n(\psi)$ profiles
for the four different $\Omega_D$ profiles with constant temperature profile $T_1$,
which are almost similar. Therefore in the limit $H_{eq}\approx E$ the potential 
profiles are almost independent of the precession frequency profiles.
One important point is that, in case of $\psi$ dependent precession frequency 
$\Omega_D(\psi)$, 
the equilibrium Hamiltonian $H_{eq}$ depends on $\psi$ in more complicated manner,
which was linear in $\psi$ for constant $\Omega_D$. Therefore the limit
$H_{eq}\approx E$ is no more a good approximation, one has to consider the expression 
of exact Hamiltonian. We will discuss this issue in the next section.
%
\section{Effect of inverse gradient length of equilibrium Hamiltonian 
$\kappa_{\Lambda}$ on TIM instability 
\label{LambdaD}}
In this section we consider the dispersion relation 
(Eq.~(\ref{Ns_res_LambdaD}-\ref{Dispersion})), which is derived from the exact 
Hamiltonian expression Eq.~(\ref{Heq_Lambda}) in sec.~\ref{Bounce_avModel}. 
In this case, according to Eq.~(\ref{KT_thr_rel_LambdaD}), the threshold value of 
$\kappa_{T}$ for TIM instability, is higher compared to the case in the limit 
$H_{eq}\approx E$.
Therefore in this section we have considered the temperature profile 
$T_1(\psi)$ with gradient $G_T =3$ and $2$ in Eq.~(\ref{T0_Profie}) which was
$0.25$ in the previous cases Sec.~\ref{compare} in the limit $H_{eq}\approx E$.
For investigating the effect of $\kappa_{\Lambda}$ on TIM instability, we 
consider precession 
frequency $\Omega_D(\psi)$ according to the theoretical expression 
Eq.~(\ref{omegaD}-\ref{omegabar}) and also the simplified expression
in the form of Eq.~(\ref{OmegaD_Psi}) with three 
different values of gradient $G_{\Omega}= 0$, $0.4$ and $0.6$. 
Fig.~\ref{FigOmegaD_Psi} presents all these $\Omega_D(\psi)$ profiles.
\begin{figure}
\includegraphics[width=\linewidth]{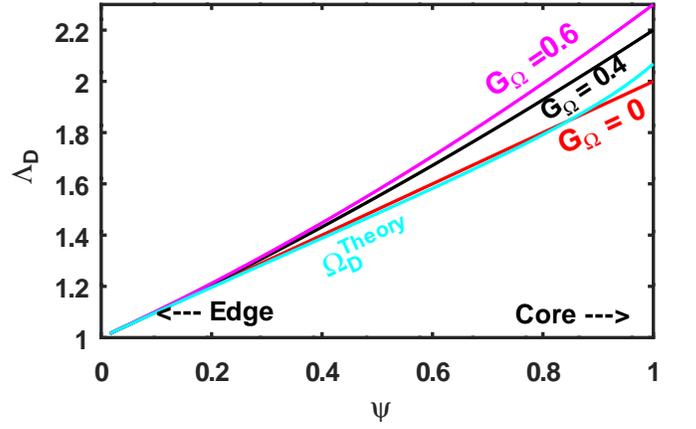}
\caption{Different profiles of $\Lambda_D(\psi) = 1+ \int \Omega_D(\psi) d\psi$, 
associated with three different $\Omega_D(\psi)$ profiles. Solid red,
black and magenta lines are associated with $G_{\Omega}= 0$, $0.4$ and $0.6$,
respectively in Eq.~(\ref{OmegaD_Psi}). The solid cyan line presents 
$\Lambda_D(\psi)$ profile for the theoretical expression of $\Omega_D(\psi)$ 
Eq.~(\ref{omegaD}-\ref{omegabar}).
\label{LambdaD_psi}}
\end{figure}
\begin{table}[h!]
\centering
\begin{tabular}{ c c c c c c c c c}
\hline
	~$\delta_{b,i}$~ &~ $\rho_{c,i}$~ &~ $C_{ad}$~ &~ $C_{pol}$~ &~ $T(\psi)$~ 
&~$T_{eq}$~ &~ $n_{eq}$~ &~ $M$~ & ~$G_{T}$~\\
\hline
~0.1~ &~ 0.03~ &~ 0.1~ &~ 0.1~ &~ $T_1$~ &~ 1~ &~ 1~ &~ 50~&~3, 2~\\
\hline
\end{tabular}
\caption{Main input parameters used for studies in sec.~\ref{LambdaD}.}
\label{table:3}
\end{table}
Considering the theoretical expression 
of $\Omega_D(r,\kappa)$ Eq.~(\ref{omegaD}-\ref{omegabar}) and the safety factor 
profile as $q(r) = 1.1 + 2 r^2$, the expression of $\Lambda_D(r,\kappa)$ 
can be written as,
\begin{eqnarray}
\begin{aligned}
	&\Lambda_D(r,\kappa) = 1+ \frac{4}{q(a)\ln{\left(\frac{q(a)}
	{q(r_{in})}\right)}} \Bigg[\Bigg.-\lambda_2(\kappa) r +
	4 \sqrt{2q(0)} \\& \lambda_1(\kappa)
	\tan^{-1}\left(\sqrt{\frac{2}{q(0)}}r\right)
	+\lambda_2(\kappa)\sqrt{\frac{1}{2}\left(q(a)-q(0)\right)}\\&
	-4 \sqrt{2q(0)}\lambda_1(\kappa)
	\tan^{-1}\left(\sqrt{\frac{q(a)}{q(0)}-1}\right)
	\Bigg.\Bigg], 
\end{aligned}
\label{LambdaD_theory}
\end{eqnarray}
where $\lambda_1(\kappa) = \left(\frac{\mathcal{E}(\kappa^2)}
{\mathcal{K}(\kappa^2)}+\kappa^2-1\right)$, and $\lambda_2(\kappa) = \left(
10 \frac{\mathcal{E}(\kappa^2)}{\mathcal{K}
(\kappa^2)} +8\kappa^2-9\right)$. Substituting the value of $r$ from 
Eq.~(\ref{Psi_r}), $\psi$ dependent $\Lambda_D(\psi,\kappa)$ can be calculated.  
This $\Lambda_{D}(\psi)$ profile Eq.~(\ref{LambdaD_theory}) with $\kappa=0$
is presented in Fig.~\ref{LambdaD_psi} with solid cyan line. The other profiles 
of $\Lambda_D(\psi)$ in Fig.~\ref{LambdaD_psi} are,
for the simplified $\Omega_D(\psi)$ profile in Eq.~(\ref{OmegaD_Psi}) with 
$G_{\Omega} = 0$, $0.4$ and $0.6$.
Therefore the $\Lambda_{D}(\psi)$ profile for the simplified $\Omega_D(\psi)$ expression
Eq.~(\ref{OmegaD_Psi}) with $G_{\Omega} = 0$, is almost similar with the 
$\Lambda_{D}(\psi)$ profile for the theoretical expression of $\Omega_D(\psi,\kappa)$
with $\kappa=0$.
The values of $\kappa_{\Lambda}= \frac{\Omega_D}{\Lambda_D}$ for all the four 
profiles at $\psi=0$ is $\kappa_{\Lambda}=1$, and then decreases with different 
rates as $\psi$ increases. 
Here first, we consider the temperature profile $T_1$ with 
gradient $G_T=3$ and study the TIM instability for all the four $\Lambda_D$ and 
$\Omega_D$ profiles. After that, for understanding the effect of temperature 
gradient on TIM instability in this new modified model with exact equilibrium
Hamiltonian, we decrease the 
temperature gradient $G_T =2$ and consider the theoretical expression of 
$\Omega_D$ as Eqs.~(\ref{omegaD}-\ref{omegabar}) and $\Lambda_D$ profile as
Eq.~(\ref{LambdaD_theory}) for $\kappa=0$.
Other essential parameters are taken from the tab.~\ref{table:3}.
The potential solution 
$\phi_n(\psi)$ is obtained by solving the differential equation 
Eq.~(\ref{QMatrix}) with the expression for $\mathcal{N}_n$ 
Eq.~(\ref{Ns_res_LambdaD}), using the spectral method. Due to high temperature 
gradient $G_T=3$ and $2$, the value of $T(\psi)$ is very high compared to previous cases,
which makes the particle equilibrium distribution $F_{eq}(\psi,\kappa,E)$ 
Eq.~(\ref{Fequilibrium_Lambda}) broaden to high $E$ value. Therefore in 
this case
we have to increase the maximum limit of $E$ as $E\in[0,45]$ for the numerical 
integration along $E$ direction, whereas for the 
previous cases with $G_T=0.25$ we considered $E\in[0,20]$.
\begin{figure}
\includegraphics[width=\linewidth]{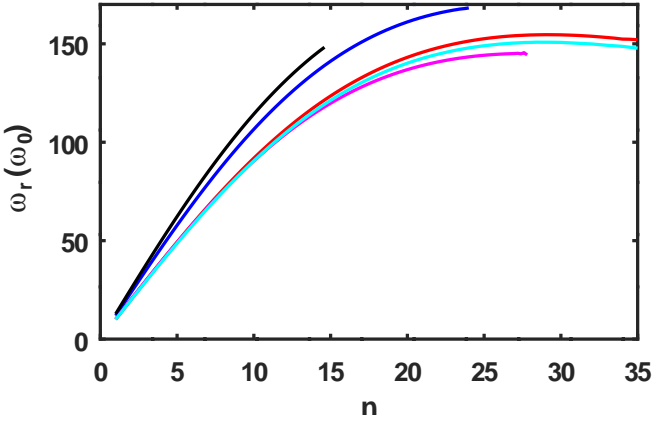}
\caption{Real part of the frequency $\omega_r$ of TIM instability for $T_1$ with
$G_T=3$ and $\Omega_D(\psi)$ profiles according to the theoretical expression
Eq.~(\ref{omegaD}-\ref{omegabar}) with $\kappa=0$ (solid cyan line), and
from the simplified expression
Eq.~(\ref{OmegaD_Psi}) with $G_{\Omega}=0$ (solid red 
line), $G_{\Omega}=0.4$ (solid blue line) and $G_{\Omega}=0.6$ (solid black line). 
The solid magenta line: for $G_T =2$ and the theoretical expression of 
	$\Omega_D(\psi)$. 
\label{Omegar_LambdaD}}
\end{figure}

Fig.~\ref{Omegar_LambdaD} presents the real part of the TIM instability for 
different mode numbers $n$ for all these five cases. 
According to Eq.~(\ref{Omegar_loc_LambdaD}), 
the case with higher $\kappa_{\Lambda}$ value has higher $\omega_r$, 
therefore the case with larger $G_{\Omega}$ has higher $\omega_r$.
In the theoretical expression of $\Omega_D$ for $\kappa=0$
the value of $\Omega_D\sim 1$ for $\psi\in[0,0.75]$ and only for $\psi\ge 0.75$,
it has values greater than unity. Therefore the $\omega_r$ profile of TIM 
instability, in this case is very close to the case with $\Omega_D = 1$ profile. 
Keeping fixed the $\Omega_D$ profile, if we decrease the $\kappa_T$ value by
decreasing $G_T=2$, the $\omega_r$ value decreases with a very small amount.
\begin{figure}
\includegraphics[width=\linewidth]{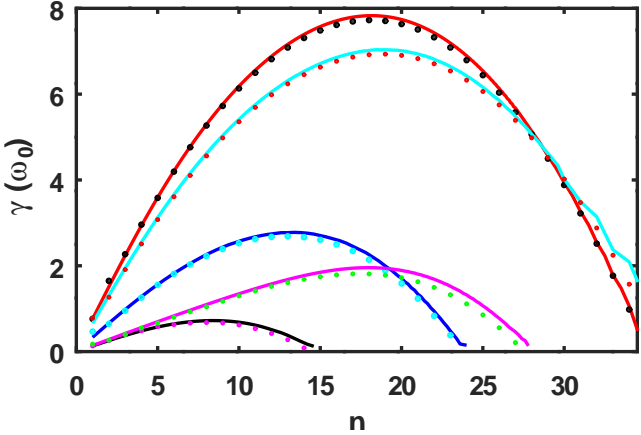}
\caption{Growth rate $\gamma$ of different modes $n$, 
$T_1$ with $G_T=3$ and different $\Omega_D(\psi)$ profiles,
        solid red, blue and black lines are for simplified $\Omega_D(\psi)$
        expression Eq.~(\ref{OmegaD_Psi}) with $G_{\Omega}=0$, $G_{\Omega}=0.4$ and
        $G_{\Omega}=0.6$ respectively.
        Solid cyan line: $\Omega_D(\psi)$ theoretical expression
        Eq.~(\ref{omegaD}-\ref{omegabar})
        with $\kappa=0$, $G_T =3$ and
       solid magenta line: for $G_T =2$ and the theoretical expression of
        $\Omega_D(\psi)$.
	The dots are the 
results from linear TERESA simulations.
\label{Growth_LambdaD}}
\end{figure}
\begin{figure}
\includegraphics[width=\linewidth]{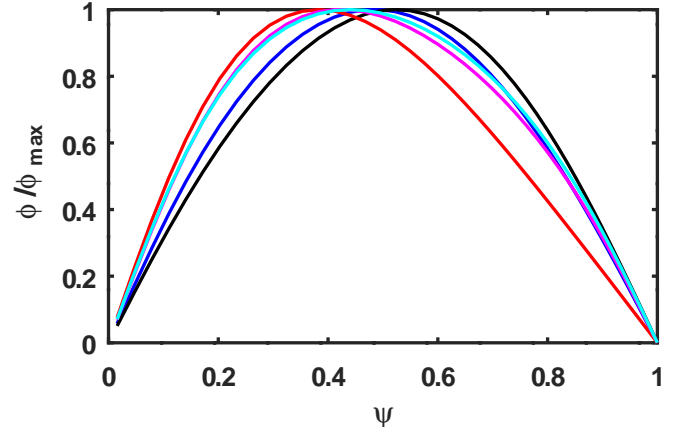}
\caption{Potential solutions $|\phi_n(\psi)|/\phi_{max}$ for mode $n=1$,
$T_1$ with $G_T=3$ and different $\Omega_D(\psi)$ profiles.
	Solid red, blue and black lines are for simplified $\Omega_D(\psi)$ 
	expression Eq.~(\ref{OmegaD_Psi}) with $G_{\Omega}=0$, $G_{\Omega}=0.4$ and 
	$G_{\Omega}=0.6$ respectively.
	Solid cyan line: $\Omega_D(\psi)$ theoretical expression 
	Eq.~(\ref{omegaD}-\ref{omegabar}) 
	with $\kappa=0$, 
       Solid magenta line: for $G_T =2$ and the theoretical expression of
        $\Omega_D(\psi)$.
\label{phi_OmegaD_LambdaD}}
\end{figure}
%
Fig.~\ref{Growth_LambdaD} presents the growth-rate $\gamma$ for different mode
numbers $n$ of TIM instability. 
Since $\kappa_{\Lambda}$ in the expression of $\mathcal{N}$ 
Eq.~(\ref{Ns_res_LambdaD}) reduces the effect of $\kappa_T$, the growth rates
in all these cases are very smaller compared to the previous cases with 
$\kappa_{\Lambda} =0$ and $\Lambda = 1$. Moreover the increase in 
$\kappa_{\Lambda}$ decreases the $\gamma$ value. Among the three $\Omega_D$ 
profiles, the cases with 
highest $G_{\Omega} =0.6$ value has higher $\kappa_{\Lambda}$ value, which gives 
the lowest growth-rate. As in the theoretical expression of 
$\Omega_D$ for $\kappa=0$ has value $\Omega_D\sim 1$ within the region 
$\psi\in [0,0.75]$ and after that it increases, therefore the growth-rate $\gamma$
in this case is almost similar to the case with $\Omega_D=1$ and have slightly 
smaller value due to $\Omega_D\ge 1$ within the region $\psi \ge 0.75$.
Therefore we can conclude that the complicated 
theoretical expression of $\Omega_D$ for $\kappa = 0$ can be simplified 
as $\Omega_D=1$ with a good accuracy for studying the TIM instability. 
If we decrease the temperature gradient $G_T=2$, for the same theoretical
$\Omega_D$ profile, the growth rate $\gamma$ decreases by a significant 
amount ($\sim70\%$) due to decrease in $\kappa_T$ value. The dots are 
the results from the linear TERESA simulations for all cases. 
Fig.~\ref{phi_OmegaD_LambdaD} presents the potential profiles $\phi_n(\psi)$ for $n=1$
and for all the five 
different cases with different $\Omega_D(\psi)$ profiles. The $\phi_n(\psi)$ solution 
are different for different $\Omega_D(\psi)$ profiles, whereas the $\phi_n(\psi)$ 
profiles for different $\Omega_D$ profiles were almost similar for the limit 
$H_{eq}\approx E$. In this new model $\phi_n(\psi)$ profiles are varying with both the 
temperature profiles and the precession frequency profiles.

Throughout this paper in the linear analysis, the nonlinear term 
$[J_{0,s}\tilde{\phi},\tilde{f}_s]_{\alpha,\psi}$ is neglected, by considering 
that $\tilde{f}_s$ is very small and the equilibrium distribution 
function $F_{eq,s}$ remains unchanged during the time evolution. However we have verified 
that, these assumptions are valid only for small amplitude potential 
$|e\phi/T| \ll 1$, and as the amplitude of the potential grows beyond a certain 
value $(e\phi/T \sim 0.1)$, due to strong nonlinear wave-particle and/or 
wave-wave interactions, 
the growthrate $\gamma(t)$ decreases and finally nonlinear saturation occurs. 
Since the nonlinear evolution of the TIM is not the focus of this study, we have not
presented those results here.
%
\section{Conclusions 
\label{Conclusions}}
In this work we proposed an alternative way to solve reduced gyro-bounce 
averaged kinetic model for trapped particle dynamics within linear limit
using a spectral method \cite{gravier:e,gravier:e13}. This method is 
computationally very fast compared to the 
semi-Lagrangian method based solver TERESA \cite{depret00:g,sarazin:y,darmet08:g}. 
Using this method we have investigated the trapped ion mode
(TIM) instability. Unlike the local linear analysis \cite{drouot14:t} 
of trapped particle mode instability, our proposed method can 
incorporate the entire profiles ($0\le\psi\le 1$) of all the essential parameters, 
and is not restricted by their local values at $\psi=0$. In that respect our
method is a {\it global-linear analysis} of trapped particle modes instability. 
Also the dependency of trapped particle drift velocity on magnetic poloidal flux 
function $\psi$, is newly incorporated in the gyro-bounce averaged trapped particle model by 
considering the exact expression of equilibrium Hamiltonian 
$H_{eq}(\psi,\kappa) = E\Lambda_D(\psi,\kappa)$ in 
quasi-neutrality equation and
gyro-bounce averaged equilibrium distribution function $F_{eq}$, which were 
previously simplified in the limit $H_{eq} \approx E$. With this new Hamiltonian,
a new quantity $\kappa_{\Lambda}$, 
that measures the inverse gradient length of equilibrium Hamiltonian, 
appears in the dispersion relation of the TIM instability.
All the previous results in the limit $H_{eq}\approx E$ can be recovered 
by substituting
$\Lambda_D= 1$ and $\kappa_{\Lambda}=0$. The quantity $\kappa_{\Lambda}$ reduces 
the effect of $\kappa_T$ and $\kappa_n$,
and as a consequence reduces the growth-rate of TIM instability. 

In tokamak plasma, trapped particle modes (resonant-branch) are driven by the resonant 
interaction 
with the precession motion of trapped particles, and these modes become unstable  
in presence of density inhomogeneity, gradient in magnetic field, and above
a critical gradient of temperature. Therefore the effect of different temperature 
profiles $T(\psi)$ and precession-frequency $\Omega_D$ on the linear TIM 
instability are investigated. The results for all the cases are compared with the
linear TERESA simulations. First we consider the model in the limit $H_{eq}\approx E$
and validate our spectral method based solver with
the TERESA simulation for the expression of gyro-bounce average operator $J_0 =1$.
The solutions from this $J_0=1$ case is used during calculation of gyro-bounce 
average of potential $J_{0}\phi$ using Pad\'e expression, for the mode number 
$n=1$ at the $1^{st}$ iteration. For studying the effect of temperature gradient 
on TIM instability, we vary the temperature profiles which are relevant to 
different regions of a tokamak plasma experiment and keep 
fixed the normalized precession frequency value $\Omega_D=1$.
Depending on the inverse temperature gradient 
length $\kappa_{T}$ value, the growth-rate of TIM instability for 
different temperature profiles is different. The profile with higher $\kappa_T$
value gives higher growth rate, and the highest growing mode also shifted
towards higher mode number. The real part 
of frequency $\omega_r - n$ profile for all the temperature profiles 
with a constant $\Omega_D$ value, follow a relation $\omega_r\sim 2 n \Omega_D$.
Then we have studied the effect of variations in precession frequency profile 
$\Omega_D$ on TIM instability in the limit $H_{eq}\approx E$. For lower value of 
$\Omega_D$ the waves with smaller
velocity make resonance with the trapped ion motion. Due to Maxwellian
energy distribution of particles, a large number of particles near lower
velocity takes parts in the instability generation mechanism, and yields 
stronger instability for smaller $\Omega_D$. Hence we get larger growth rate
for smaller $\Omega_D$. 
The effect of pitch-angle dependency of trapped particles on the TIM instability 
is investigated by considering the entire $\kappa$ dependent profile of 
$\Omega_D$. 
Since for a constant magnetic shear $s_0=0.8$, for most of the 
$\kappa$ value $\Omega_D(\kappa)>1$ except near the separatrix 
$0.85\le\kappa\le1$, the growth-rate of TIM instability is smaller compared to 
$\Omega_D = 1$ case. The $\psi$ dependency of the precession frequency 
$\Omega_D(\psi)$ on TIM instability in the limit $H_{eq}\approx E$, is also 
investigated for a simplified linearly increasing $\Omega_D$ profile.
Since for this profile the values of precession frequency $\Omega_D>1$ for all 
the $\psi>0$ location, the growth-rate of the TIM instability is smaller than the
case with $\Omega_D = 1$. In the limit $H_{eq}\approx E$, the potential profiles
are almost independent of different $\Omega_D(\psi,\kappa)$ profiles.

The effect of $\psi$ dependency of $\Omega_D$ on the TIM 
instability in the newly modified model with exact equilibrium Hamiltonian expression 
is investigated by considering a 
theoretical expression of $\Omega_D$
and a simplified linearly increasing function of $\Omega_D$. For $\kappa=0$
the theoretical expression of $\Omega_D$ has value $\Omega_D\sim 1$ within the
region $0\le\psi\le 0.75$, and after that it increases. 
The growth-rate $\gamma$ of TIM instability for the particular 
theoretical $\Omega_D$ profile is almost similar to the $\Omega_D=1$ case, and
due to $\Omega_D\ge 1$ with in the region $0.75\le\psi\le 1$, it is slightly smaller
than $\Omega_D=1$ case.
With increase of the slope $G_{\Omega}$ in the simplified linearly 
increasing $\Omega_D$ profile, which increases $\kappa_{\Lambda}$ 
value, the growth-rate of different modes $n$ for the instability decreases.
Moreover after the new modification, the decrease in slope $G_T$ of the 
temperature profile (decrease $\kappa_T$ value) for a fixed $\Omega_D$ profile,
decreases the growth rate, which is consistent with the results in the limit 
$H_{eq}\approx E$. In this newly modified model the potential solutions are different for
different $\Omega_D$ profiles.
%
\section*{Acknowledgements}
This work was funded by the Agence Nationale de la Recherche for the Project GRANUL 
(ANR-19-CE30-0005). This work was granted access to the HPC resources of EXPLOR 
(Project No. 2017M4XXX0251), and of CINECA MARCONI under Project FUA35 GSNTITE. 
This work has been carried out within the framework of the EUROfusion Consortium, 
funded by the European Union via the Euratom Research and Training Programme 
(Grant Agreement No 101052200-EUROfusion). Views and opinions expressed are however 
those of the author(s) only and do not necessarily reflect those of the European Union 
or the European Commission. Neither the European Union nor the European Commission 
can be held responsible for them. We are grateful to the anonymous reviewers
for their constructive remarks.
\appendix
\section{Linear analysis within the limit $H_{eq}\approx E$ \label{Appd_HeqE}}
%
The global linear analysis for the case with the limit $H_{eq}\approx E$ can be 
derived in the similar way, as discussed in Sec.~\ref{Global_linear}
Here the normalized equilibrium distribution $F_{eq,s}$ is 
independent of ($\alpha,\kappa, t$), and has the form of a two dimensional 
Maxwellian energy distribution function
\begin{eqnarray}
	F_{eq,s}(\psi,E) = \frac{n_{s}(\psi)}{T_{s}^{3/2}(\psi)} 
	\exp\left(-\frac{E}{T_{s}(\psi)}\right). 
\label{Fequilibrium_HE}
\end{eqnarray}
After substituting $F_{eq,s}$ from Eq.(\ref{Fequilibrium_HE}),
$\tilde{f}_s = \sum_{n,\omega} f_{s,n,\omega}(\psi,E,\kappa) 
\exp\{i(n\alpha -\omega t)\}$
and $\tilde{\phi} = \sum_{n,\omega} \phi_{n,\omega}(\psi) 
\exp\{i(n\alpha -\omega t)\}$, in 
Eq.~(\ref{linear-vlasov}), the solution of Vlasov equation in Fourier space become,
\begin{eqnarray}
\begin{aligned}
	f_{s,n,\omega}(\psi,E,\kappa) =& \frac{n\left[ \kappa_n(\psi) + \kappa_{T}(\psi)
	\left(\frac{E}{T_s(\psi)}-\frac{3}{2}\right)\right]}
	{Z_s^{-1}n\Omega_{D}(\kappa) E-\omega}&& \\& 
	\left\{J_{0,n,s}\phi_{n,\omega}(\psi)\right\} F_{eq,s}(\psi,E),
\end{aligned}
\label{linearfSol_HE}
\end{eqnarray}
In this case the elementary volume in phase-space can be written as 
$d^3v = 4\pi\sqrt{2}m^{-3/2}\sqrt{E}dE \frac{d\lambda}{4\Omega_{D}}$.
Using this volume element $d^3v$ and $f_{n,\omega}$ from 
Eq.~(\ref{linearfSol_HE}) the expression of $\mathcal{N}_s$ in 
Eq.~(\ref{quasi-Nutr2}) can be written 
(in the limit of a constant {\it pitch-angle}) as,
\begin{eqnarray}
\begin{aligned}
	\mathcal{N}_{n,s}(\psi) = \frac{1}{n_{eq}}\int_{0}^{\infty}
	J_{0,n,s}\Bigg[\Bigg.\frac{\kappa_n(\psi) + \kappa_{T}(\psi)
        \left(\frac{E}{T_s(\psi)}-\frac{3}{2}\right)}
	{Z_{s}^{-1}\Omega_{D} (E-\chi_s)} \\
        \left\{J_{0,n,s}\phi_{n,\omega}(\psi)\right\}
	\frac{n_s(\psi)}{T_s^{3/2}(\psi)} 
        \exp\left(-\frac{E}{T_s(\psi)}\right)
	\Bigg.\Bigg]\sqrt{E} dE, 
\end{aligned}
\label{Ns_res_HE}
\end{eqnarray}
where $\chi_s = \frac{\omega}{nZ_s^{-1}\Omega_D}$. 
The expression of $C_n$
Eq.~(\ref{CnDefine}) remains unaltered.
Therefore the dispersion 
relation becomes,
\begin{eqnarray}
	C_n\phi_{n,\omega} = \mathcal{N}^*_{n,i} \phi_{n,\omega} 
	-\mathcal{N}^*_{n,e} \phi_{n,\omega} 
\label{Dispersion_HE}
\end{eqnarray}
where $\mathcal{N}^*_{n,s} = \frac{\mathcal{N}_{n,s}}{\phi_{n,\omega}}$.
The final form of the 2nd-order 
differential equation for $\phi_n(\psi)$ Eq.~(\ref{QMatrix}) remains unchanged, 
except $\mathcal{N}^*_n$ inside $Q_n(\psi)$ Eq.~(\ref{QMatrix1sp}), where 
$\mathcal{N}$ is given by Eq.~(\ref{Ns_res_HE}).
Comparing the expressions of $\mathcal{N}$ from Eq.~(\ref{Ns_res_HE}) and 
Eq.~(\ref{Ns_res_LambdaD}),
the main difference in the dispersion relation arises due to the absence of
the term $-\frac{E\Lambda_D}{T} \kappa_{\Lambda}$ in $\mathcal{N}$ in the limit
$H_{eq} \approx E$, which reduces the effective contribution from 
$\kappa_n$ and $\kappa_T$ in case of exact Hamiltonian. Therefore the growth-rate of 
the TIM instability in the limit $H_{eq} \approx E$, is significantly higher
compared to the case with exact Hamiltonian. 

As discussed in Sec.~\ref{local_linear} the local linear stability analysis of this 
reduced gyro-bounce averaged model in the limit $H_{eq}\approx E$ can be 
performed and the threshold frequency value of TIM
instability can be derived as \cite{drouot14:t}:
\begin{eqnarray}
	\omega_r =\frac{ \left(\frac{3}{2} \kappa_T -\kappa_n \right)}{\kappa_T}n \Omega_D T_{0}, 
\label{Loc_linHE}
\end{eqnarray}
where $T_0$ is the temperature at $\psi=0$. 
The threshold value of $\kappa_{T}$ for the TIM instability can be written as
\begin{eqnarray}
        \kappa_{T,th} = \frac{C_n\Omega_{D0}}{\int_0^{\infty} J_{0,n}^2 
        \exp(-\xi)\sqrt{\xi} d\xi} 
\label{KT_thrHE}
\end{eqnarray}
where $\xi = \frac{E}{T}$.

\end{document}